\documentclass[reqno,12pt]{amsart}
\usepackage{epsfig}\usepackage{amsfonts,amsmath,amssymb,mathrsfs,verbatim,amsthm,amscd}
\usepackage{cite}
\usepackage{epsf}\usepackage{amssymb} \usepackage{cite}
\usepackage{amsthm}
\usepackage{amscd}
\usepackage{amsfonts}
\usepackage{amsmath}
\usepackage{hyperref}
\usepackage{mathtools} %,dutchcal} % математика и символы
\usepackage{mathtext} % русские буквы в математических выражениях
\usepackage{bm} % bold под наклоном

\numberwithin{equation}{section}

\renewcommand{\phi}{\varphi}

\renewcommand{\epsilon}{\varepsilon}

\DeclareMathOperator{\sh}{sh}
\DeclareMathOperator{\ch}{ch}

\renewcommand{\imath}{\mathrm{i}}

\let\Re\relax
\let\Im\relax
\DeclareMathOperator{\Re}{Re}
\DeclareMathOperator{\Im}{Im}
\newcommand{\be}{\begin{equation}}
\newcommand{\ee}{\end{equation}}
\newcommand{\bea}{\begin{eqnarray}}
\newcommand{\eea}{\end{eqnarray}}

\def\stackreb#1#2{\ \mathrel{\mathop{#1}\limits_{#2}}}

\usepackage{xcolor}

\begin{document}

\title[Complex rational Ruijsenaars model. The two-particle case]
{Complex rational Ruijsenaars model. \\ The two-particle case}

\author{N. \,M. Belousov}%

\address{N. B.: Beijing Institute of Mathematical Sciences and Applications,
Huairou district, Beijing, 101408, China}

\author{G. \,A. Sarkissian}%

\address{G. S.: Laboratory of Theoretical Physics,
JINR, Dubna, Moscow region, 141980 Russia and
Yerevan Physics Institute, Alikhanian Br. 2, 0036\, Yerevan, Armenia}

\author{V. \,P. Spiridonov}%

\address{V. S.: Laboratory of Theoretical Physics,
JINR, Dubna, Moscow region, 141980 Russia and
National Research University Higher School of Economics, Moscow, Russia}

%\vskip 1em

\begin{abstract} \noindent
We consider a complex rational degeneration of the hyperbolic Ruijsenaars model
emerging in the limit $\omega_1+\omega_2\to 0$ (or $b\to \imath$ in $2d$ CFT) and
investigate in detail the two-particle case. Corresponding wave functions are
described by complex hypergeometric functions in the Mellin-Barnes representation.
Their dual integral representation   and reflection symmetry in the coupling constant are established.
Besides, a complex limit of the hyperbolic Baxter $Q$-operators is considered.
Another complex degeneration of the hyperbolic Ruijsenaars model is
obtained by taking a special $\omega_1-\omega_2\to 0$ (or $b\to 1$) limit.
Additionally, two new degenerations to the complex Calogero-Sutherland type models
are described.
\end{abstract}

\maketitle

\tableofcontents

%\vspace*{0.5em}

\section{Introduction}

The many-body models initially considered by Calogero and Sutherland are among the
most popular completely integrable systems \cite{H,OlPe}.
Their ubiquitous nature is reflected in the number of applications they have found in
theoretical physics and mathematics. Notably they were introduced from the physical considerations
in quantum mechanics and solid states physics. Later they emerged in the theory of solitons (e.g.,
in the dynamics of poles of rational KP equation  solutions), quantum field theory and statistical
mechanics mirrored by the mathematical applications in the representation theory, theory of multiple
orthogonal polynomials, symplectic and algebraic geometry, and so on.

Theory of integrable systems is deeply related to the theory of special functions. However, the
notions of integrable and exactly solvable models differ from each other. The former one
requires existence of sufficiently many integrals of motion basically without restrictions on
the type of functions the dynamics of systems is described by. The latter one is demanding that
the functions involved in the model solution should have some closed form representations.
This includes at least an explicit series representation of the functions or a definite integral
representation (the most popular options which are assumed to produce asymptotic estimates
of the functions). In exceptional situations an infinite product representation of functions
is available when the full set of divisor points is known. It often happens that integrable
systems appear to be exactly solvable. If not, it serves as a challenge to the theory
of special functions to make them of that type.

An important generalization of the standard Calogero-Sutherland
models was suggested by Ruijsenaars in a long series of papers, which can be considered in the physical
setting as a relativistic extension of non-relativistic systems of particles. The most structurally
complicated systems correspond to the Hamiltonians described by finite-difference operators
with the elliptic function coefficients (potentials) \cite{R87,vD0}. The hyperbolic Ruijsenaars
system \cite{R11} is obtained by degenerating the double (quasi)periodic functions to the functions
with one period (corresponding to the standard degeneration of the modular parameter $\Im(\tau)\to +\infty$).
And here lies the dividing line between the integrable and exactly solvable models: the Ruijsenaars hyperbolic model was fully solved in recent papers \cite{halik,halnas,HR3} and \cite{Belousov:2023qgn, Belousov:2023bisp, Belousov:2023ort,Belousov:2023znq}, while the elliptic one still remains not tractable in the closed
form from this point of view. Only special explicit
solutions are known for the van Diejen model \cite{vD0}, a multiparameter extension of the elliptic
Ruijsenaars system, which are expressed in terms of the elliptic hypergeometric functions
\cite{AN,spi:cs} or Bethe ansatz type solutions \cite{Ch}.

The elliptic hypergeometric equation \cite{spi:cs,spi:essays} is the most general known
second order finite-difference equation having solutions in the closed form. Its coefficients are elliptic functions depending
on 8 parameters. This equation serves as a laboratory of investigating
translation invariant two-particle systems or one-particle systems when the translational  symmetry
is absent. Its degeneration to the hyperbolic and trigonometric levels was considered in  \cite{BRS}
(see also \cite{Derkachov:2021thp,Apresyan:2022erh}).
Reduction of the solutions to the standard rational level was described in \cite{rai:limits}. In \cite{Sarkissian:2020ipg,SaSpJPA} a new reduction of the equation and its solutions to the level
of complex hypergeometric functions was discovered. The uniformity of such a limit is justified
by the asymptotic considerations given in \cite{rai:limits}.   The reduction of hypergeometric-type special functions led to a new rational degeneration of the hyperbolic Ruijsenaars and van Diejen systems in \cite{SaSpJPA}.

In the present paper we limit our consideration to the two-body hyperbolic Ruijsenaars model
and investigate its different degeneration limits. The Hamiltonian
eigenvalue problem in the hyperbolic case is related to the $q$-hypergeometric equation
in the regime $|q|=1$ considered in \cite{NU,Rbanff}. The corresponding wave function has
found applications in the Liouville theory
(two-dimensional conformal field theory), where it describes $S$-generator of one point
conformal blocks \cite{T2003} (see also \cite{AS} and references therein).
For a more recent analysis of the relation to modular group $SL(2,\mathbb{Z})$ see \cite{DFKKSS}.
Also it is related to the generator of $S$-duality in four-dimensional $N=4$ gauge field theory
and partition functions in supersymmetric electrodynamics on squashed three-sphere
\cite{HLP} (for a relation to the hyperbolic Ruijsenaars system see also \cite{BKK}).
Note the general hierarchical structure of these objects: the elliptic
hypergeometric integrals --- the top special functions of hypergeometric type --- describe
superconformal indices of four-dimensional $N=1$ supersymmetric models and their
degeneration to the hyperbolic level yields supersymmetric partition functions of
three-dimensional theories on the squashed sphere \cite{DSV}.

More specifically, we solve the eigenvalue problem for the degeneration
of the hyperbolic two-particle Ruijsenaars model Hamiltonian to the complex hypergeometric level
in the special limit $\omega_1+\omega_2\to 0$ (or $b=\sqrt{\omega_1/\omega_2}\to \imath$ in $2d$ CFT).
We describe symmetries of the corresponding wave functions and derive the dual spectral problems.
Together with that we describe related Baxter $Q$-operators and their commutativity. In order to reach these results we apply the new degeneration limit
for hyperbolic hypergeometric functions discovered in our recent paper \cite{BSS}.
We also use some old technique for getting symmetries of involved hyperbolic integrals
described in \cite{Kashaev}.

Besides, we describe a special singular $\omega_1\to\omega_2$ (or $b\to 1$) limit from the hyperbolic  two-particle
Ruijsenaars model to a model resembling the rational Ruijsenaars system. However, we postpone
consideration of the corresponding wave functions to a later work. Additionally, we
present two degenerations of the hyperbolic system
to new differential  Calogero-Sutherland type systems. The first one  is
related to the spectral problem considered in \cite{MN} on the basis of complex hypergeometric equation.
The second one is also related to the complex hypergeometric equation, but now it is qualitatively
different from the first case. The physical interpretation of the
latter two models is not clear yet, although they look like the models of particles moving on a cylinder.

\section{Degenerations of the two-particle hyperbolic Ruijsenaars model}

The hyperbolic Ruijsenaars model with two particles is described by two Hamiltonians associated with
the shifts of the physical coordinate along two quasiperiods $\omega_1$ and $\omega_2$.
In the center of mass frame one has \cite{R11}
\begin{align}\label{Hh}
	& H_h = \frac{ \sh \frac{\pi}{\omega_2} (x - \imath g) }{ \sh \frac{\pi}{\omega_2} x } \, e^{- \imath \omega_1 \partial_x} + \frac{ \sh \frac{\pi}{\omega_2} (x + \imath g) }{ \sh \frac{\pi}{\omega_2} x } \, e^{\imath \omega_1 \partial_x},
\\[6pt] \label{Hh2}
	& H_h' = \frac{ \sh \frac{\pi}{\omega_1} (x - \imath g) }{ \sh \frac{\pi}{\omega_1} x } \, e^{- \imath \omega_2 \partial_x} + \frac{ \sh \frac{\pi}{\omega_1} (x + \imath g) }{ \sh \frac{\pi}{\omega_1} x } \, e^{\imath \omega_2 \partial_x},
\end{align}
where $e^{\imath \omega_k \partial_x}f(x):=f(x+\imath \omega_k)$.
Because the coefficients in front of the shift operators are periodic functions, these
Hamiltonians commute with each other, $[H_h,H_h']=0$.
It appears that the eigenvalue problems for these operators have solutions analytic in $x$,
see Section~\ref{sec:hyp-func}.

Define the scalar product
\begin{equation}
\langle \varphi|\psi\rangle_h:=\int_{\mathbb{R}}\mu(x)\,
\overline{\varphi(x)}\psi(x)\,\frac{dx}{\sqrt{\omega_1\omega_2}},
\quad \mu(x)=\frac{\gamma^{(2)}(g\pm \imath x;\mathbf{\omega})}
{\gamma^{(2)}(\pm \imath x;\mathbf{\omega})},
\label{SP_h}\end{equation}
where $\gamma^{(2)}(x;\mathbf{\omega})$ is the hyperbolic gamma function or the
Faddeev modular dilogarithm whose properties are described in Appendix~\ref{app:hgamma}.   Here and in what follows we use shorthand notation
\begin{equation}
	f(x \pm y) = f(x + y) f(x - y)
\end{equation}
for various functions $f$.

For $g\in \mathbb{R}$ and either $\omega_1, \omega_2\in \mathbb{R}$ or $\bar \omega_1=\omega_2$
the weight function is   nonnegative, $\mu(x)\geq 0$.
  Under additional assumptions $g, \Re \omega_1, \Re \omega_2 > 0$ (notice that the case $g = 0$ is trivial) two sequences of the weight function poles
\begin{align}
	x_{\mathrm{poles}} = \pm \imath (g + m_1 \omega_1 + m_2 \omega_2), \qquad m_1, m_2 \in \mathbb{Z}_{\geq 0}
\end{align}
are separated by the integration contour $\mathbb{R}$.
In this case it is not difficult to check that for real quasiperiods both Hamiltonians are formally self-adjoint
$H_h^\dag=H_h$ and $(H_h')^\dag=H_h'$, i.e.
$$
\langle \varphi|H_h\psi\rangle_h=\langle H_h\varphi|\psi\rangle_h, \qquad
\langle \varphi|H_h'\psi\rangle_h=\langle H_h'\varphi|\psi\rangle_h,
$$
  assuming $\varphi, \psi$ are analytic and decay sufficiently fast in the strip
  $| \Im  x| \leq \max(\Re \omega_1, \Re \omega_2)$. Moreover, these operators can be made
  bona fide self-adjoint under additional assumption $g \leq \omega_1 + \omega_2$~\cite{R11}.
  For complex conjugated quasiperiods $\bar\omega_1=\omega_2$ one has $H_h^\dag=H'_h$.

There is the second choice of parameter $g$ for which one can propose a nonnegative measure~\cite{DFKKSS}.
Namely, let us fix
\begin{equation} \label{g-unitarity2}
	g=\tfrac12(\omega_1+\omega_2)+\imath a,\quad a\in\mathbb{R},
\end{equation}
and define the scalar product (cf. \cite{Ponsot:2000mt})
\begin{equation}
\langle \varphi|\psi\rangle_{h2}:=\int_{\mathbb{R}}\mu(x)\,
\overline{\varphi(x)}\psi(x)\,\frac{dx}{\sqrt{\omega_1\omega_2}},
\quad \mu(x)=\frac{1}{\gamma^{(2)}(\pm \imath x;\mathbf{\omega})}
=4\sh \frac{\pi x}{\omega_1}\sh \frac{\pi x}{\omega_2}.
\label{SP_h2}\end{equation}
Then, for $\omega_1,\, \omega_2 > 0$ we have
again $H_h^\dag=H_h$ and $(H_h')^\dag=H_h'$ (in this case ``halves'' of these
Hamiltonians are separately formally self-adjoint), whereas for $\bar \omega_1=\omega_2$ the operators are adjoint to each other $H_h^\dag=H'_h$.

  In total, we have two measure functions and four unitarity regimes of parameters $g, \omega_1, \omega_2$. The corresponding unitary transforms diagonalizing the Hamiltonians are treated in uniform manner in the paper~\cite{BK}.

Let us describe six qualitatively different degenerations of the hyperbolic Ruijsenaars model.
Two of them are well known, the third was discovered recently in \cite{SaSpJPA} and the other three are new ones.

\subsection{Rational Ruijsenaars model}

Let us fix $g, \omega_1, \omega_2 \in \mathbb{R}$, rescale the variables as
\begin{align}
\qquad g = \omega_1 b , \qquad x \to \omega_1 x
\end{align}
and take the limit $\omega_1\to 0$ while keeping $\omega_2$ fixed.
The parameter $b$ introduced here should not be mixed up with the $b$-variable used in $2d$ CFT.
In this case
\begin{align}
	\frac{ \sh \frac{\pi\omega_1}{\omega_2} (x - \imath b) }{ \sh \frac{\pi\omega_1}{\omega_2} x } \underset{ \omega_1 \to 0}{=} \frac{x - \imath b}{x},
\end{align}
and the first Hamiltonian $H_h$ reduces to
\begin{align}
	H_r = \frac{x - \imath b}{x} e^{- \imath \partial_x} + \frac{x + \imath b}{x} e^{ \imath \partial_x}.
\end{align}
The second Hamiltonian $H_h'$ does not have definite limit for $\omega_1 \to 0$.
Still, there is an example \cite{SaSpJPA} when for a fixed eigenvalue of a similar (but more complicated)
operator one can get some nontrivial relations in this limit.

Due to formula~\eqref{r_real} the limiting scalar product \eqref{SP_h} takes the following form
(up to the diverging factor $(2\pi)^{2g}(\omega_1/\omega_2)^{2g+1/2}$)
\begin{equation}
\langle \varphi|\psi\rangle_r:=\int_{\mathbb{R}}\mu(x)\,
\overline{\varphi(x)}\psi(x)\, dx, \quad \mu(x)=\frac{\Gamma(b\pm \imath x)}
{\Gamma(\pm \imath x)},
\label{SP_r}\end{equation}
  so that under the assumption $b > 0$ (which ensures that the gamma function poles are separated by the integration contour $\mathbb{R}$) we have   $\langle \varphi|H_r\psi\rangle_r=\langle H_r\varphi|\psi\rangle_r$. This scalar product is defined
up to the multiplication of the measure by some periodic function $\mu_0(x+\imath )=\mu_0(x)$.   Let us remark that in \cite{Belousov:2023sat} a slightly different parametrization $g = \omega_1 b + \omega_2$ was considered, so that in the resulting measure $\mu(x)$ the product $\Gamma(b\pm \imath x)$ was replaced
by $1/\Gamma(1-b\pm \imath x)$, which corresponds to the choice
$\mu_0(x)=\sin\pi( b\pm \imath x)/\pi^2$.

This is the standard rational degeneration of the hyperbolic model considered already by Ruijsenaars \cite{Rbanff}.
Equivalently, this reduction can be done by taking the limit $\omega_2 \to \infty$
with fixed $\omega_1,x, g$. Note that the second scalar product
\eqref{SP_h2} does not have definite form in this limit.

\subsection{Complex rational Ruijsenaars model} \label{sec:cr-deg}

Here we describe a reduction of the hyperbolic Ruijsenaars model to the complex rational model discovered
in \cite{SaSpJPA}. Let us fix the following complex conjugated periods
\begin{align}\label{lim_cc}
	\omega_1 = \imath + \delta, \qquad \omega_2 = - \imath + \delta,
\end{align}
so that $\omega_1+\omega_2=2\delta$ and $\omega_1\omega_2=1+\delta^2$. Furthermore, consider the parametrization
\begin{align}
	x =\sqrt{\omega_1\omega_2}( k + u \delta), \qquad g = \imath \sqrt{\omega_1\omega_2}(\ell + h \delta), \qquad k, \ell \in \mathbb{Z},
\end{align}
and take the limit $\delta\to 0^+$. Then
\begin{align}
	\frac{ \sh \frac{\pi}{\omega_2} (x - \imath g) }{ \sh \frac{\pi}{\omega_2} x } \underset{\delta \to 0^+}{=} (-1)^\ell \; \frac{ z + b }{ z },
\end{align}
where
\begin{align}
	z = \frac{k + \imath u}{2}, \qquad b = \frac{\ell + \imath h}{2}.
\end{align}
Besides,
$$	
	x - \imath \omega_1 = \sqrt{\omega_1 \omega_2} \bigl( k + 1 + (u - \imath) \delta + O(\delta^2) \bigr),
$$
so that for the function $\Phi(x)$ which has the limit
\begin{align}
	\Phi(x) \underset{\delta \to 0^+}{\propto} \Psi(k, u),
\end{align}
with a possible diverging proportionality factor, we have
\begin{align}
	e^{- \imath \omega_1 \partial_x} \Phi(x):= \Phi(x - \imath \omega_1) \underset{\delta \to 0^+}{\propto} \Psi(k + 1, u - \imath) =: e^{\partial_z} \Psi(k, u),
\end{align}
i.e. we can write formally $e^{\partial_z}:=e^{\partial_k}e^{-\imath \partial_u}$.
Thus, modulo the sign $(-1)^\ell$ the first Hamiltonian $H_h$ \eqref{Hh} reduces to the operator
\begin{align}\label{Hcr}
	H_{cr} = \frac{z + b}{z} \, e^{\partial_z} + \frac{z - b}{z} \, e^{ - \partial_z}.
\end{align}
Note that the operator $e^{\partial_z}$ acts on the functions of $z$ as a simple shift operator,
$e^{\partial_z}f(z)=f(z+1)$. Therefore one could think that the eigenvalue problem for
the Hamiltonian \eqref{Hcr}, $H_{cr}\Psi=\lambda\Psi$,  is an analytic difference equation yielding as a solution an analytical function of $z$, $\Psi=\Psi(z)$, defined up to a periodic function factor $c(z+1)=c(z)$.
However, this is not so, the solution is
not obliged to be of such a form --- this phenomenon was discovered in \cite{SaSpJPA} on the
example of available degenerations of the elliptic hypergeometric equation.

Indeed, denote in an analogous way
\begin{align}
	z' = \frac{-k + \imath u}{2}, \qquad b' = \frac{-\ell + \imath h}{2}.
\end{align}
Then the second Hamiltonian $H'_h$ \eqref{Hh2} reduces up to the sign factor $(-1)^\ell$ to
\begin{align}
	H'_{cr} = \frac{z' + b'}{z'} \, e^{\partial_{z'}} + \frac{z' - b'}{z'} \, e^{ -\partial_{z'}},
\label{H'_cr}\end{align}
where $e^{\partial_{z'}} \Psi(k, u) := \Psi(k -1, u - \imath)$, i.e. formally
$e^{\partial_{z'}}:=e^{-\partial_k}e^{-\imath \partial_u}$. Again, the reduced Hamiltonians commute
with each other, $[H_{cr},H'_{cr}]=0$. Their joint eigenvalue problems restrict the form of eigenfunctions,
but still do not define them uniquely (up to the multiplication by a constant).

The limiting transition $\delta\to 0^+$ from the hyperbolic to complex hypergeometric integrals was considered
in \cite{Sarkissian:2020ipg}. We describe it below in detail on the example of degeneration
of the wave function \eqref{int-to-sum}.
Applying it to the measure in \eqref{SP_h}, one can see that infinitely many poles approach
the integration contour and asymptotically the integral gets replaced by an infinite sum of simpler
integrals near each pole
$$
\int_{\mathbb{R}}\mu(x)\,\frac{dx}{\sqrt{\omega_1\omega_2}}
\to e^{\pi\imath \ell}(4\pi\delta)^{2h}\sum_{k\in\mathbb{Z}}\int_{\mathbb{R}}   \mu(k,u)\, du,
\quad \mu(k,u)=\frac{\bm{\Gamma}(b\pm z|b'\pm z')}{\bm{\Gamma}(\pm z|\pm z')},
$$
where $\bm{\Gamma}(z|z')$ is the complex gamma function described in the Appendix \ref{app:hgamma} below
and we use the notation $\bm{\Gamma}(b\pm z|b'\pm z')=\bm{\Gamma}(b+ z|b'+ z')\bm{\Gamma}(b-z|b'-z')$.
We drop the diverging factor and define the scalar product for eigenfunctions of $H_{cr}$
and $H_{cr}'$ as follows
\begin{equation}
\langle \varphi|\psi\rangle_{cr}:= \sum_{k\in\mathbb{Z}}\int_{\mathbb{R}}   \mu(k,u)\,
\overline{\varphi(k,u)}\psi(k,u)\, du.
\label{SP_cr}\end{equation}

If we replace in \eqref{SP_cr} the function $\overline{\varphi(k,u)}$ by $\varphi(k,u)$,
then both Hamiltonians become symmetric with respect to the   resulting bilinear form,
$$
\langle \varphi|H_{cr}\psi\rangle_{cr}=\langle H_{cr}\varphi|\psi\rangle_{cr},\qquad
\langle\varphi|H_{cr}'\psi\rangle_{cr}=\langle H_{cr}'\varphi|\psi\rangle_{cr},
$$
  assuming $\Im  h < 0$. Note that under this assumption two sequences of the weight function poles
\begin{align}
	u_{\mathrm{poles}}  = \mp h  \pm \imath (|\ell \pm k| + 2m), \qquad m \in \mathbb{Z}_{\geq 0},
\end{align}
are separated by the integration contour $\mathbb{R}$.

Since the variable $u$ is real, $z'$ is related to the complex conjugation of the
complex variable $z$, $z'=-\bar z$. Consequently there emerge the following
relations between the formal operators $e^{\partial_z}$ and $e^{\partial_{z'}}$
$$
{\overline {e^{\partial_z}}}=e^{\partial_{\bar z}}= e^{\partial_k}e^{+\imath \partial_u}=e^{-\partial_{z'}}.
$$
As a result, one finds the conjugation of the Hamiltonian $H_{cr}$ with respect to the
scalar product \eqref{SP_cr} to be
\begin{equation}
H_{cr}^\dag=\frac{\bar z + \bar b}{\bar z} \, e^{\partial_{\bar z}} + \frac{\bar z
- \bar b}{\bar z} \, e^{ -\partial_{\bar z}}=
\frac{z' -\bar b}{z'} \, e^{-\partial_{z'}} + \frac{z' +\bar b}{z'} \, e^{\partial_{z'}}.
\label{H_crHC}\end{equation}
Comparing this expression with \eqref{H'_cr}, we see that $H_{cr}^\dag=H'_{cr}$,
provided $\bar b=b'$, which is equivalent to the restrictions
\begin{equation}
\ell=0, \qquad   b=\frac{\imath h}{2} > 0.
\label{unitarity}\end{equation}
Under these conditions the weight function is   nonnegative, $\mu(k,u)\geq 0$,
$$
\mu(k,u)
=(-1)^{b-b'}\frac{\bm{\Gamma}(b+z|b'+ z')}{\bm{\Gamma}(z|z')}
\frac{\bm{\Gamma}(b'-z'|b-z)}{\bm{\Gamma}(-z'|- z)}
=\bigg|\frac{\bm{\Gamma}(b+z|b-\bar z)}{\bm{\Gamma}(z|-\bar z)}\bigg|^2
=\big|z\bm{\Gamma}(b+z|b-\bar z)\big|^2.
$$
The combinations $H_{cr}+H_{cr}'$ and $\imath (H_{cr}-H_{cr}')$ are formally self-adjoint under the conditions \eqref{unitarity}.   Note that these conditions are consistent with the assumption $g > 0$ in the hyperbolic model, since we used the ansatz $g = \imath \sqrt{1 + \delta^2}(\ell + h \delta)$ with $\ell \in \mathbb{Z}$.

As to the second scalar product \eqref{SP_h2}, it reduces up to the factor $16\pi^2\delta^2$ to the form
\begin{equation}
\langle \varphi|\psi\rangle_{cr2}:=
\sum_{k\in\mathbb{Z}}\int_{\mathbb{R}}\mu(k,u)\,
\overline{\varphi(k,u)}\psi(k,u)\, du,
\quad \mu(k,u)= |z|^2 \geq 0.
\label{SP_cr2}\end{equation}
In this case we have
\begin{equation}
\langle \varphi|H_{cr}\psi\rangle_{cr2}=
\sum_{k\in\mathbb{Z}}\int_{\mathbb{R}}\,|z|^2\,\overline{\varphi(k,u)}
\Bigl( \frac{z + b}{z} \, e^{\partial_z} + \frac{z - b}{z} \, e^{ - \partial_z} \Bigr)\psi(k,u)\,du
 =\langle H_{cr}^\dag \varphi|\psi\rangle_{cr2},
\label{SP_cr2conj}\end{equation}
where
$$
H_{cr}^\dag =\frac{\bar z -1+\bar b}{\bar z} \, e^{-\partial_{\bar z}}
+ \frac{\bar z +1- \bar b}{\bar z} \, e^{\partial_{\bar z}}.
$$
Changing the parametrization of $g$ to
$$
g=\tfrac12(\omega_1+\omega_2)+\imath a=\delta+\imath a=\imath\sqrt{\omega_1\omega_2}(\ell+\delta(-\imath+h)),
$$
we obtain
\begin{equation}
b=\tfrac12 +\tfrac12(\ell+\imath h), \quad b'=\tfrac12 +\tfrac12(-\ell+\imath h), \qquad   \ell \in \mathbb{Z}, \; h \in \mathbb{R}.
\label{b2}\end{equation}
As a result, $1-\bar b=b'$ and for this scalar product one has $H_{cr}^\dag=H'_{cr}$.
Below we will derive eigenfunctions of the Hamiltonians $H_{cr}$ and $H_{cr}'$
for general parameter~$b$, but for physical
interpretation of them as some quantum mechanical wave functions it will be necessary to impose
either the constraints \eqref{unitarity} or \eqref{b2}.

\subsection{Complementary complex rational Ruijsenaars model}

Now we consider a special singular limit for hyperbolic hypergeometric integrals when
$\omega_1\to\omega_2$, which was discovered in \cite{Sarkissian:2020ipg}.
Consider complex conjugated periods
\begin{align}
	\omega_1 = 1+\imath\delta, \qquad \omega_2 = 1- \imath\delta,
\end{align}
and take
\begin{align}
	x =\imath\sqrt{\omega_1\omega_2}(- k + u \delta), \qquad g =\sqrt{\omega_1\omega_2}( \ell - h \delta), \qquad k, \ell \in \mathbb{Z}.
\end{align}
Then,
\begin{align}
	\frac{ \sh \frac{\pi}{\omega_2} (x - \imath g) }{ \sh \frac{\pi}{\omega_2} x } \underset{\delta \to 0^+}{=} (-1)^\ell \; \frac{ z + b }{ z },
\end{align}
where we have the same variables $z$ and $b$ as before. The degeneration of the elliptic
hypergeometric equation in this limit was considered in \cite{SaSpJPA}. Existence
of the related analogue of the rational Ruijsenaars model follows from this analysis, but it
was not considered in that paper, which we do now.

Consider the corresponding limit of Hamiltonian operators. The change of the shift operators
in $H_h$ is the same as in the complex hypergeometric case,
$e^{- \imath \omega_1 \partial_x} \to e^{\partial_z}$.
As a result, we see that the first Hamiltonian has exactly the same form as before
\begin{align}\label{Hccr}
	H_{ccr} = \frac{z + b}{z} \, e^{\partial_z} + \frac{z - b}{z} \, e^{ - \partial_z}= H_{cr}.
\end{align}
However, the second Hamiltonian gets a different form.
Up to the sign factor $(-1)^\ell$, we have
\begin{align}\label{Hccr'}
	H_{ccr}' = \frac{z' + b'}{z'} \, e^{-\partial_{z'}} + \frac{z' - b'}{z'} \, e^{ \partial_{z'}} \neq H'_{cr}.
\end{align}
Although we have ``small'' difference in the operators, the solutions of the corresponding
eigenvalue problems are substantially different due to
qualitatively distinct degenerations of the hyperbolic gamma function in these limits
\cite{Sarkissian:2020ipg}. An explicit example of such a difference is described in \cite{SaSpJPA}.

Direct application of the limit~\eqref{gamma-poh-limit} to the weight function \eqref{SP_h} yields
\begin{eqnarray}\nonumber && \makebox[4em]{}
\mu(x)\stackreb{=}{\delta\to 0^+}
e^{\pi\imath \ell}   \, (4\pi\delta)^{2\ell}\, \mu(k,u),\quad
\\  &&
\mu(k,u)=\frac{\Gamma(b\pm z)\Gamma(1\pm z')}{\Gamma(\pm z)\Gamma(b'+1\pm z')}
=\frac{(1-\frac{\ell-\imath h\pm(k-\imath u)}{2})_{\ell\pm k-1}}
{(1\mp \frac{k-\imath u}{2})_{\pm k-1}}.
\label{WFccr}\end{eqnarray}
As we see, there are poles depending on the integer $k$ and, following the form of the measure in
the complex Ruijsenaars model, we propose to consider the following scalar product
\begin{equation}
\langle \varphi|\psi\rangle_{ccr}:=\sum_{k\in\mathbb{Z}}\int_{C_k}   \mu(k,u)\,
\overline{\varphi(k,u)}\psi(k,u)\, du,
\label{SP_ccr}\end{equation}
where $C_k$ are some contours separating measure poles coming from different sign
choices of $\pm k$ in the Pochhammer symbols.
If we take $h\in\mathbb{R}$, then $b'=-\bar b$ for arbitrary $\ell$ and $h$.
As a result, we obtain the formal Hermitian conjugation rule $H_{ccr}^\dag=H'_{ccr}$, i.e.
$\langle \varphi|H_{ccr}\psi\rangle_{ccr}=\langle H'_{ccr}\varphi|\psi\rangle_{ccr}$
without further restrictions on $b$.

Since $z'=-\bar z$, the taken weight function $\mu(k,u)$ can be nonnegative
only for $b=-b'=1$, which reduces it to $|z|^4$. One can multiply it by some  function
which is periodic with respect to the shifts $z\to z+1$ and $z'\to z'+1$ to remove this restriction.
For instance, we can take
$$
\tilde\mu(k,u)=\frac{\sin\pi(\pm z')}{\sin\pi(b'\pm z')}\mu(k,u)
=\frac{\Gamma(b\pm z)\Gamma(-b'\pm z')}{\Gamma(\pm z)\Gamma(\pm z')},
$$
where we used the inversion relation $\Gamma(x)\Gamma(1-x)=\pi/\sin\pi x$.
This weight function is nonnegative, $\tilde\mu(k,u)=|\Gamma(b\pm z)/\Gamma(\pm z)|^2\geq 0$.
However, now there are infinitely many poles for a fixed summation index $k$, i.e. we
have a transcendental function instead of the rational one. Moreover, this $\tilde \mu(k,u)$
cannot be obtained as a limit from the original measure \eqref{SP_h}.

As to the second weight function \eqref{SP_h2}, it is reduced similarly to the previous case
to $\mu(k,u)=|z|^2$ up to the multiplicative factor  $-16\pi^2\delta^2$, which can be dropped out.
We parametrize as before $b=\tfrac12(\ell+\imath h)$ and find the same $H_{ccr}^\dag=H_{cr}^\dag$.
Since $z'=-\bar z$, we have $H_{ccr}^\dag=H'_{ccr}$ provided $\bar b-1=b'$,
  which is equivalent to $\ell = 1$, $\Re h = 0$. In other words,
\begin{equation}
	b = \tfrac{1}{2} + r, \qquad b' = -\tfrac{1}{2} + r, \qquad r \in \mathbb{R}.
\end{equation}
This is consistent with the hyperbolic model unitarity condition~\eqref{g-unitarity2}
with $a\propto \delta\Im  h$.

In the present paper we do not consider solutions of the eigenvalue problem in the complimentary
complex rational case postponing it to the future.

\subsection{Hyperbolic Calogero-Sutherland model}

Consider the standard degeneration limit from the hyperbolic Ruijsenaars model to the well known
Calogero-Sutherland system. Let us renormalize the coupling constant $g = b \omega_1,$ take
fixed $x$, $b$, normalize the quasiperiod $\omega_2=1$, and consider the limit $\omega_1\to 0$.
Then the coefficients in the eigenvalue equation for the Hamiltonian $H$ take the form
\begin{align}
	\frac{ \sh \pi(x - \imath b \omega_1) }{ \sh \pi x }
= 1 - \imath \pi b \omega_1 \coth\pi x - \frac{\pi^2b^2 \omega_1^2}{2} + O(\omega_1^3).
\end{align}
Besides,
\begin{align}
	\Phi(x - \imath \omega_1) = \Phi(x) - \imath  \omega_1 \partial_x \Phi(x)
- \frac{ \omega_1^2 }{2} \partial_x^2 \Phi(x) + O(\omega_1^3),
\end{align}
or in other words
\begin{align}
	e^{- \imath \omega_1 \partial_x} = 1 - \imath  \omega_1 \partial_x
- \frac{ \omega_1^2 }{2} \partial_x^2  + O(\omega_1^3).
\end{align}
From these expansions we find
\begin{align}
	\frac{1}{\omega_1^2} (H_h - 2) \underset{\omega_1 \to 0^+}{=} - \partial_x^2 - 2\pi b \coth \pi x
\, \partial_x -(\pi b)^2 =: H_{CS}.
\end{align}

Consider how the measure \eqref{SP_h} is degenerated in this  limit.
Using formula \eqref{r_real_rat}, we find
$$
\mu(x)\stackreb{=}{\omega_1\to 0} |2\sh\pi x|^{2b}.
$$
  Clearly, for $b \in\mathbb{R}$ it is nonnegative. The scalar product does not change and we have formally self-adjoint Hamiltonian $H_{CS}^\dag=H_{CS}$   (e.g. on the space of smooth compactly supported functions on $\mathbb{R}\setminus \{0\}$). Let us note that after the change of variable $w = -\sh^2 \pi x$ the Hamiltonian $H_{CS}$ becomes a hypergeometric operator, which can be made bona fide self-adjoint under the assumption $b > 0$~\cite{N}.

Renormalizing the wave function $\psi(x)\to | 2\sh\pi x|^{-b}\psi(x)$ we arrive at the Hamiltonian
$$
\tilde H_{CS}=|2\sh \pi x|^{b} \, H_{CS} \, |2\sh\pi x|^{-b}= - \partial_x^2 + \pi^2\frac{b(b-1)}{\sh^2\pi x},
$$
which is evidently Hermitian for the constant weight function. Similarly to the standard rational
Ruijsenaars model, the second scalar product \eqref{SP_h2} does not have definite limit in this case as well.

\subsection{Complex hyperbolic Calogero-Sutherland model} \label{sec:comp-cs}

  Now we describe a new degeneration to a complex version of the Calogero-Sutherland model. As discussed in Section~\ref{sec:dual},
  it is ``dual'' to the complex rational Ruijsenaars model from Section~\ref{sec:cr-deg}.

Consider the following complex conjugated periods
\begin{align}
	\omega_1 = \imath + \delta, \qquad \omega_2 = - \imath + \delta
\end{align}
and set
\begin{align}
	x =\sqrt{\omega_1\omega_2}(N + \beta), \qquad
g = \imath\sqrt{\omega_1\omega_2} (\ell + h \delta), \qquad N,\ell \in\mathbb{Z},\quad \beta, h\in\mathbb{C}.
\end{align}
Take the limits $N \to \pm \infty$ and $\delta \to 0^+$ simultaneously in such a way that
\begin{align}
	N \delta \to \alpha \in \mathbb{R},
\end{align}
where $\alpha$ is some fixed number. Denoting
\begin{align}
	z = \alpha + \imath \beta, \qquad b = \frac{\ell + \imath h}{2},
\end{align}
we have
\begin{eqnarray}\nonumber &&\makebox[-1em]{}
\sh \frac{\pi}{\omega_2} (x \mp \imath g)   =(-1)^{N+\ell}\sh \pi(z+\delta (A\pm B)+O(\delta^3))
\\ \label{AB} &&
=(-1)^{N+\ell} \bigl(1+ \delta \pi (A\pm B)\coth \pi z+\tfrac12 \delta^2 \pi^2 (A\pm B)^2 \bigr) \sh \pi z+O(\delta^3),
\end{eqnarray}
where
$$
A:=\beta -\tfrac{\imath}{2} \alpha-\tfrac12 z \delta,\quad B:= 2b +\delta(h-\tfrac{\imath}{2}\ell).
$$
The coordinate $x$ is replaced by the variables $\alpha$ and $\beta$. Therefore
we replace the Hamiltonian eigenfunctions $\Phi(x)$  by the functions of the latter variables
\begin{align}
	\Phi(x) \underset{\substack{\; \delta \to 0^+,\, N \to\infty}}{\propto} \Psi(\alpha, \beta),
\end{align}
where we assume that the diverging numerical factors (if any) are dropped out.
Because we have
\begin{align}
	x - \imath \omega_1 = \sqrt{\omega_1 \omega_2} \biggl( \frac{\alpha + \delta}{\delta} + (\beta - \imath \delta) + O(\delta^2) \biggr),
\end{align}
we can write
\begin{align}
	e^{- \imath \omega_1 \partial_x} \Phi(x) := \Phi(x - \imath \omega_1)
\underset{\substack{\; \delta \to 0^+,\, N \to\infty}}{\propto} \Psi(\alpha + \delta, \beta - \imath \delta) = e^{\delta (\partial_\alpha - \imath \partial_\beta)} \Psi(\alpha, \beta).
\end{align}
Since $\partial_\alpha - \imath \partial_\beta=2\partial_z$, in other words we have
\begin{align}
	e^{- \imath \omega_1 \partial_x} \to e^{2\delta \partial_z}= 1 +2 \delta \partial_z +2\delta^2 \partial_z^2 + O(\delta^3).
\end{align}
From the above expansions we obtain
\begin{align}
\frac{1}{4\delta^2}(2 - (-1)^\ell H_h) \underset{\substack{\; \delta \to 0^+
\\[3pt] N \to\infty}}{=} - \partial_z^2 - 2\pi b  \coth(\pi z) \,\partial_z - (\pi b)^2 =: H_c.
\label{Hc}\end{align}
This is a complex version of the hyperbolic Calogero-Sutherland model. For $\beta\in \mathbb{R}$ it
corresponds to a system with the configuration space being a cylinder with the coordinates
$\alpha\in\mathbb{R}$ and $\beta\in \bigl[-\frac{1}{2}, \frac{1}{2} \bigr]$.

The second Hamiltonian \eqref{Hh2} also has a similar limit in terms of the variables
\begin{align}
	\bar z = \alpha - \imath \beta, \qquad b' = \frac{- \ell + \imath h}{2}.
\end{align}
Namely, we have again the expansion \eqref{AB} with $z \to \bar{z}$ and the coefficients
$$
A:=\beta +\tfrac{\imath}{2} \alpha-\tfrac12 \bar z \delta,\quad B:= 2b' +\delta(h+\tfrac{\imath}{2}\ell).
$$
Analogously,
$$
e^{- \imath \omega_2 \partial_x} \to e^{-\delta (\partial_\alpha + \imath \partial_\beta)}=
 e^{-2\delta \partial_{\bar z}}= 1 -2 \delta \partial_{\bar z} +2\delta^2 \partial_{\bar z}^2 + O(\delta^3).
$$
Substitution of these expansions in \eqref{Hh2} yields
\begin{align}
\frac{1}{4\delta^2}(2 - (-1)^\ell H'_h) \underset{\substack{\; \delta \to 0^+
\\[3pt]
N \to\infty}}{=} - \partial_{\bar z}^2 - 2\pi b'  \coth(\pi \bar z) \,\partial_{\bar z} - (\pi b')^2 =: H_c'.
\label{Hc'}\end{align}

After the change of variable $w = - \sh^2 \pi z$  eigenvalue problems for the operators
$H_{c}$ and $H'_{c}$ with $b = b' \in \mathbb{R}$
yield a system of two complex conjugated hypergeometric equations coinciding
with the one considered in \cite{MN} for the parameter identifications $a_{MN}=b/2,\, b_{MN}=1/2 +b/2$.
Note also that for the choice $b=b'=1/2$ Hamiltonians $H_c$ and $H_c'$ describe Casimir
operators of the Lorentz group in the sector of zero magnetic quantum numbers
\cite{SK}, where $z$ variable is the complexified rotation angle
with $\alpha$ corresponding to the boost and $\beta$ to the ordinary rotation parameters.

In order to find the limiting form  of the weight function we use the results of our
recent paper \cite{BSS}. Applying formula~\eqref{lim2'} we find
\begin{align}
	\frac{\gamma^{(2)}(g \pm \imath x)}{\gamma^{(2)}(\pm \imath x)} \underset{\substack{\delta \to 0 \\[2pt] N \to \infty}}{=}  \bigl( 2\sh \pi z \bigr)^{2b} \, \bigl(2 \sh \pi \bar z \bigr)^{2b'},
\end{align}
which is a nonnegative function for $b = b' \in \mathbb{R}$.   The scalar product in the limit is given by a two-dimensional integral over the cylinder
\begin{equation}
\langle \varphi|\psi\rangle_c:=
\int_{-\frac{1}{2}}^{\frac{1}{2}} d\beta \int_{-\infty}^\infty d\alpha \, \mu(z) \,
\overline{\varphi(\alpha, \beta)}\,\psi(\alpha, \beta),\quad
\mu(z)= \bigl| 2\sh \pi z \bigr|^{4b},
\label{SP_c}\end{equation}
which coincides after the change of variable $w = - \sh^2 \pi z$ with the scalar product
considered in~\cite{MN} for the corresponding parameters $a_{MN}=b/2,\, b_{MN}=1/2 +b/2$.
  The precise mechanism describing how the univariate integrals with hyperbolic gamma functions degenerate into
  such two-dimensional integrals will be considered in~\cite{BSS2}.

  On the space of smooth compactly supported functions $\psi(\alpha, \beta)$ on $\mathbb{R} \times \mathbb{S} \setminus \{0, 0\}$
  the Hamiltonians are formally adjoint $H_c^\dag=H'_c$. In the two-parameter model of~\cite{MN}
analogues of the operator combinations  $H_c + H_c'$ and $\imath(H_c - H_c')$ were shown to have self-adjoint
extensions under the condition $0<a_{MN}+b_{MN}<2$, which reduces in our case to  $-\frac{1}{2} < b < \frac{3}{2}$.

Renormalizing the wave functions $\psi \to |2\sh\pi z|^{-2b}\psi$ we arrive at the Hamiltonians
\be
\tilde H_c= - \partial_z^2 + \pi^2\frac{b(b-1)}{\sh^2\pi z},\qquad
\tilde H_c'=\tilde H_c^\dag= - \partial_{\bar z}^2 + \pi^2\frac{b(b-1)}{\sh^2\pi \bar z}.
\label{Hcd}\end{equation}
Their Hermitian combination
$$
\tilde{H}_c+\tilde{H}'_c=-\tfrac12\partial_\alpha^2+\tfrac12\partial_\beta^2+\pi^2b(b-1)\Bigl(\frac{1}{\sh^2\pi z}
+\frac{1}{\sh^2\pi \bar z}\Bigr)
$$
looks like a Hamiltonian describing a particle moving
on the cylinder with some specific potential, but the kinetic energy of the $\beta$-degree of
freedom is negative which is unusual.

The second weight function~\eqref{SP_h2} is reduced in the taken limit to the expression
$\mu(z)=| 2\sh \pi z|^2$. Then it is easy to find that
$$
H_{c}^\dag= - \partial_{\bar z}^2 +2\pi (\bar b-1)\coth (\pi\bar z) \partial_{\bar z} -\pi^2(1-\bar b)^2.
$$
For  $\bar b-1=-b'$, or equivalently
\begin{equation}
	b=\tfrac12 +\tfrac12(\ell+\imath h), \qquad b'=\tfrac12 +\tfrac12(-\ell+\imath h), \qquad \ell \in \mathbb{Z}, \; h \in \mathbb{R},
\end{equation}
we have $H_{c}^\dag=H'_{c}$. The system with the second scalar
product was investigated in the recent paper \cite{spin2}, where it emerged in the consideration
of open non-compact spin chain. Note also that in the special case $b=b'=1/2$ the measure \eqref{SP_c} coincides with the second one and it becomes precisely
the left-invariant Haar measure of the Lorentz group described in \cite{SK}.

\subsection{Complementary complex Calogero-Sutherland model}

The last limiting model we consider corresponds to the following choice of the hyperbolic
Ruijsenaars system variables. Take  quasiperiods
\begin{align}
	\omega_1 = 1+\imath\delta, \qquad \omega_2 = 1- \imath\delta,
\end{align}
and parametrize
$$
x=\imath \sqrt{\omega_1 \omega_2} (N + \beta ), \quad
g = \sqrt{\omega_1 \omega_2}(\ell + h \delta).
$$
As before, take the limits $N \to \pm \infty$, $\delta \to 0^+$ with
$N \delta \to\alpha \in \mathbb{R}$  and denote $\bar z = \alpha - \imath \beta$,
$b' = \frac{-\ell + \imath h}{2}$.
Then we have again the expansion \eqref{AB} with $z \to - \bar{z}$ and the coefficients
$$
A:=-\beta -\tfrac{\imath}{2} \alpha+\tfrac12 \bar z \delta,\quad B:= -2b' +\delta(h+\tfrac{\imath}{2}\ell).
$$
Similar to the previous case we replace the Hamiltonian eigenfunctions $\Phi(x)$  by the functions
$\Psi(\alpha, \beta)$ up to some numerical (possibly diverging) factor. From the relation
\begin{align}
	x - \imath \omega_1 = \imath \sqrt{\omega_1 \omega_2} \biggl( \frac{\alpha - \delta}{\delta} + \beta - \imath \delta + O(\delta^2) \biggr),
\end{align}
we obtain
\begin{align}
		e^{- \imath \omega_1 \partial_x} \Phi(x) := \Phi(x - \imath \omega_1)
\underset{\substack{\; \delta \to 0^+,\, N \to\infty}}{\propto} \Psi(\alpha - \delta, \beta - \imath \delta) = e^{-\delta (\partial_\alpha + \imath \partial_\beta)} \Psi(\alpha, \beta).
\end{align}
Recall that $\partial_\alpha + \imath \partial_\beta=2\partial_{\bar z}$. Therefore,
\begin{align}
	e^{- \imath \omega_1 \partial_x} \to e^{-2\delta \partial_{\bar z}}= 1 -2 \delta \partial_{\bar z}
+2\delta^2 \partial_{\bar z}^2 + O(\delta^3).
\end{align}
As a result, we obtain
\begin{align}
\frac{1}{4\delta^2}(2 - (-1)^\ell H_h) \underset{\substack{\; \delta \to 0^+
\\[3pt] N \to\infty}}{=} - \partial_{\bar z}^2 + 2\pi b'  \coth(\pi \bar{z}) \,\partial_{\bar z}  - (\pi b')^2 =: H_{cc}.
\label{H_cc}\end{align}

Analogously, for the second hyperbolic Hamiltonian~\eqref{Hh2} we have the same expansion \eqref{AB} with
$$
A:=\beta -\tfrac{\imath}{2} \alpha-\tfrac12 z \delta,\quad B:= -2b - \delta(h-\tfrac{\imath}{2}\ell)
$$
and
$$
e^{- \imath \omega_2 \partial_x} \to e^{-\delta (\partial_\alpha - \imath \partial_\beta)}=
 e^{-2\delta \partial_{ z}}.
$$
As a result,
\begin{align}
\frac{1}{4\delta^2}(2 - (-1)^\ell H'_h) \underset{\substack{\; \delta \to 0^+
\\[3pt]
N \to\infty}}{=} - \partial_z^2 - 2\pi b  \coth(\pi  z) \,\partial_z - (\pi b)^2=:H_{cc}'.
\label{H_cc'}\end{align}
This is again a complex version of the hyperbolic Calogero-Sutherland model. However, it differs from
the previous case by the sign of coupling constant $b'$, that is $H_{cc}' = H_c$, but $H_{cc} \neq H'_c$, see~\eqref{Hc'}.

In similarity with the previous complex Calogero-Sutherland model, we take as a scalar product the
following two-dimensional integral over the cylinder
\begin{eqnarray}\nonumber  &&
\langle \varphi|\psi\rangle_{cc}:=
\int_{-\frac{1}{2}}^{\frac{1}{2}} d\beta \int_{-\infty}^\infty d\alpha \, \mu(z) \,
\overline{\varphi(\alpha, \beta)} \, \psi(\alpha, \beta),\quad
\\ && \makebox[1em]{}
\mu(z)= \bigl( 2\sh \pi z \bigr)^{2b} \, \bigl(2 \sh \pi \bar z \bigr)^{-2b'}
= \big|(2\sh \pi z)^{2b}\big|^2.
\label{SP_cc}\end{eqnarray}
  In the last equality we take $h\in\mathbb{R}$ which results in the complex conjugation rule
$\bar b=-b'$ leading to the nonnegative weight function.
Because for this scalar product
$$
H^\dag_{cc}=H'_{cc}=\bar{H}_{cc},
$$
we conclude that eigenvalue problems for the Hamiltonians
$H_{cc}$ and $H'_{cc}$ yield the hypergeometric equation
in the variable $w=-\sh^2 \pi z$ and its complex conjugate. These equations depend on one discrete $\ell \in \mathbb{Z}$
and one real $h\in \mathbb{R}$ parameters, which substantially differs from the system considered in \cite{MN}.
Renormalizing the wave functions $\psi \to |(2\sh\pi z)^b|^{-2}\psi$ we obtain the Hamiltonians
\begin{equation}
\tilde H_{cc}= - \partial_{\bar z}^2 + \pi^2\frac{\bar b(\bar b-1)}{\sh^2\pi\bar z} ,\qquad
\tilde H_{cc}'=\tilde H_{cc}^\dag= - \partial_{z}^2 + \pi^2\frac{b(b-1)}{\sh^2\pi z}.
\label{Hccd}\end{equation}
For $\bar b=b=\ell/2$ (i.e. $h=0$) these operators coincide with the complex Hamiltonians \eqref{Hcd}.
The constraint $-\tfrac12<b<\tfrac{3}{2}$ yields three possible choices: $\ell=0,2$ leading to the free systems
$b(b-1)=0$ and $\ell=1$ corresponding to the standard Lorentz group Casimir operators \cite{SK}.

Application of the limit $\omega_1 \to \omega_2$ to the second scalar product \eqref{SP_h2} results in the same
weight function, as in the previous complex degeneration, $\mu(z)\propto|\sh\pi z|^2$.   Assuming $\bar b -1 = b'$, or equivalently
\begin{align}
	b = \tfrac{1}{2} + r, \qquad b' = -\tfrac{1}{2} + r, \qquad r \in \mathbb{R},
\end{align}
we obtain formally adjoint operators
$H_{cc}^\dag = H'_{cc}$.

At the moment, we do not know the geometric and group-theoretical interpretation of the general emerging model.
We only stress that eigenfunctions of the Hamiltonians in this case are not described in general
by the complex hypergeometric functions of the type considered in \cite{Neretin:2019xze}.

\section{Hyperbolic wave function} \label{sec:hyp-func}

In order to investigate eigenvalue problems for the $\omega_1+\omega_2\to 0$ degenerated Hamiltonians,
we derive a master difference equation and its solution independently on the unitarity conditions
associated with the nonnegativity of the corresponding measures. Some limits of the elliptic hypergeometric equation were considered
in \cite{BRS,Derkachov:2021thp,Apresyan:2022erh}. Our starting point is
the following reduced difference equation at the hyperbolic level derived there
\be\label{secdif}
{\mathcal D}(\underline{\mu},\underline{\nu})(U(\mu_2+\omega_2,\mu_3-\omega_2)-U(u))+(\mu_2\leftrightarrow \mu_3)+U(u)=0,
\ee
where
\bea
{\mathcal D}(\underline{\mu},\underline{\nu})={\sin{{\pi\over\omega_1}}(\mu_2-\mu_4-\omega_2)\sin{{\pi\over \omega_1}}(\mu_4-\mu_2)
\over \sin{{\pi\over \omega_1}}(\mu_2-\mu_3)\sin{{\pi\over \omega_1}}(\mu_3-\mu_2-\omega_2)}
\prod_{k=1}^4{\sin{{\pi\over \omega_1}}(\mu_3+\nu_k-\omega_2)\over \sin{{\pi\over \omega_1}}(\mu_4+\nu_k)},
\eea
\be
 U(\underline{\mu},\underline{\nu})={J_h(\mu,\nu)\over \gamma^{(2)}(\mu_2- \mu_4, \mu_3- \mu_4;\mathbf{\omega})}
\ee
and
\be\label{jhh}
J_h(\underline{\mu},\underline{\nu})=\int_{-\textup{i}\infty}^{\textup{i}\infty}\prod_{j=1}^4
\gamma^{(2)}(\mu_j- z,\nu_j+ z;\mathbf{\omega}){dz\over \textup{i}\sqrt{\omega_1\omega_2}}.
\ee
Here the parameters $\mu_j$, $\nu_j$ satisfy the balancing condition
\be\label{ball}
\sum_{j=1}^4(\nu_j+\mu_j)=2Q:=2(\omega_1+\omega_2)
\ee
and we use the notation $\gamma^{(2)}(a,\ldots,b;\mathbf{\omega})=
\gamma^{(2)}(a;\mathbf{\omega})\cdots\gamma^{(2)}(b;\mathbf{\omega})$.
Now we take in (\ref{secdif}) the limit
\be\label{111}
\mu_2\to -\textup{i}\infty \quad {\rm and}\quad  \nu_2=2Q-\sum_{j=1,3,4}(\nu_j+\mu_j)-\mu_2\to \textup{i}\infty.
\ee
Using relations \eqref{gamasym} and \eqref{gamasym2}, we obtain the equation
\be\label{eq5}
{\mathcal B}_1{\mathcal A}_1+{\mathcal B}_2{\mathcal A}_{-1}=({\mathcal B}_1+{\mathcal B}_2-1){\mathcal A}_0,
\ee
where
\be\nonumber
{\mathcal B}_1=e^{{\pi\textup{i} \over \omega_1}(\mu_3-\mu_4-\omega_2)}\prod_{k=1,3,4}{\sin{{\pi\over \omega_1}}(\mu_3+\nu_k-\omega_2)\over \sin{{\pi\over \omega_1}}(\mu_4+\nu_k)}
\ee
and
\be\nonumber
{\mathcal B}_2=e^{{\pi\textup{i} \over \omega_1}(\mu_3-\mu_1)}{\sin{{\pi\over\omega_1}}(\mu_3-\mu_4-\omega_2)\sin{{\pi\over \omega_1}}(\mu_4-\mu_3)
\sin{{\pi\over \omega_1}}(2Q-\omega_2-\sum_{k=1,3,4}(\mu_k+\nu_k))\over \sin{{\pi\over \omega_1}}(\mu_4+\nu_1)
\sin{{\pi\over \omega_1}}(\mu_4+\nu_3)\sin{{\pi\over \omega_1}}(\mu_4+\nu_4)},
\ee
\bea\nonumber &&
{\mathcal A}_a={e^{{\pi\textup{i} \over \omega_1} a \mu_4}\over \gamma^{(2)}(\mu_3- a \omega_2-\mu_4)}
\int_{-\textup{i}\infty}^{\textup{i}\infty}
e^{{\pi\textup{i} z \over \omega_1\omega_2}(\sum_{k=1,3,4}(\mu_k+\nu_k)-Q- a \omega_2)}
\nonumber\\  && \makebox[2em]{} \times
\prod_{k=1,3,4}\gamma^{(2)}(\nu_k+z;\mathbf{\omega})\, \gamma^{(2)}(\mu_1-z,\mu_3- a \omega_2-z,
\mu_4-z;\mathbf{\omega}){dz\over \textup{i}\sqrt{\omega_1\omega_2}}.
\eea
Next we take in (\ref{eq5}) the limit
$\mu_3\to \textup{i}\infty$ and $\nu_3\to -\textup{i}\infty$ with $\mu_3+\nu_3=y$ kept fixed.
In this way we obtain the equation
\be\label{555}
{\mathcal D}_1{\mathcal E}_1+{\mathcal D}_2{\mathcal E}_{-1}=({\mathcal D}_1+{\mathcal D}_2-1){\mathcal E}_0,
\ee
\be\label{d11}
{\mathcal D}_1={1\over 2\textup{i}}e^{{\pi\textup{i} \over \omega_1}(-y-2\mu_4-\nu_1-\nu_4+\omega_2)}{\sin{{\pi\over \omega_1}}(y-\omega_2)\over \sin{{\pi\over \omega_1}}(\mu_4+\nu_1)\sin{{\pi\over \omega_1}}(\mu_4+\nu_4)},
\ee
\be
{\mathcal D}_2=-{1\over 2\textup{i}}e^{{\pi\textup{i} \over \omega_1}(-y-\mu_1+\mu_4+\omega_2)}{
\sin{{\pi\over \omega_1}}(2Q-\omega_2-\sum_{k=1,4}(\mu_k+\nu_k)-y)\over \sin{{\pi\over \omega_1}}(\mu_4+\nu_1)
\sin{{\pi\over \omega_1}}(\mu_4+\nu_4)},
\ee
\bea\label{betainter2} \nonumber  &&
{\mathcal E}_a= e^{{2\pi\textup{i} \over \omega_1} a \mu_4}\int_{-\textup{i}\infty}^{\textup{i}\infty}
e^{{\pi\textup{i}z \over \omega_1\omega_2}(\sum_{k=1,4}(\mu_k+\nu_k)+2y-2Q-2a \omega_2)}
\\ && \makebox[3em]{} \times
\prod_{k=1,4}\gamma^{(2)}(\nu_k+z,\mu_k-z;\mathbf{\omega}){dz\over \textup{i}\sqrt{\omega_1\omega_2}}.
\eea
Let us relabel the variables $\mu_4\to \mu_2,\, \nu_4\to \nu_2$, introduce the parameter $x$
\be\label{aprima}
x=y-Q+{1\over 2}\sum_{k=1,2}(\mu_k+\nu_k)
\ee
and the function
\be\label{newf}
F(\underline{\nu};\underline{\mu};x)
={1\over \gamma^{(2)}(g;\mathbf{\omega})}
\int_{-\textup{i}\infty}^{\textup{i}\infty}e^{{2\pi\textup{i}xz \over \omega_1\omega_2}}\prod_{k=1,2}\gamma^{(2)}(\nu_k+z,\mu_k-z;\mathbf{\omega})
{dz\over \textup{i}\sqrt{\omega_1\omega_2}}.
\ee
Now equation (\ref{555}) can be rewritten in the form
\bea\nonumber &&
e^{-{\pi\textup{i} \over \omega_1}C}\sin{\pi\over \omega_1}(x+B)F(\underline{\nu};\underline{\mu};x+\omega_2)
+e^{{\pi\textup{i} \over \omega_1}C}\sin{\pi\over \omega_1}(x-B)F(\underline{\nu};\underline{\mu};x-\omega_2)
\\ && \makebox[2em]{}
=\textup{i}\Big(e^{-{\pi\textup{i} \over \omega_1}x}\cos{\pi\over \omega_1}(\mu_2-\mu_1)-e^{{\pi\textup{i} \over \omega_1}x}\cos{\pi\over \omega_1}(\nu_2-\nu_1)\Big)
F(\underline{\nu};\underline{\mu};x),
 \label{888}\eea
where
\be\label{bd}
B={1\over 2}\sum_{k=1,2}(\mu_k+\nu_k)\quad {\rm and} \quad C={1\over 2}\sum_{k=1,2}(\mu_k-\nu_k).
\ee

Using the pentagon identity \cite{Faddeev:2000if,Ponsot:2000mt}
\bea\label{pentafkv3}  \makebox[-2em]{}
&&  \int_{-\textup{i}\infty}^{\textup{i}\infty}
e^{{2 \pi\textup{i} \over \omega_1\omega_2} y(z+(\nu-\mu)/2)}
 \gamma^{(2)}(\pm y+(Q-\nu-\mu)/2;\mathbf{\omega})
{dy\over \textup{i}\sqrt{\omega_1\omega_2}}
\nonumber \\ && \makebox[2em]{}
= \frac{\gamma^{(2)}\left(z+\nu,-z +\mu;\mathbf{\omega}\right)}{\gamma^{(2)}(\nu+\mu;\mathbf{\omega})},
\eea
one can find \cite{Kashaev} (see also \cite{BSS}) the dual form of the function \eqref{newf}:
\bea\nonumber  && \makebox[-2em]{}
F(\nu_1,\nu_2;\mu_1,\mu_2;x)=e^{-{ \pi\textup{i} \over 2\omega_1\omega_2} x(\beta_1+\beta_2)}\gamma^{(2)}(\alpha_1,\alpha_2;\mathbf{\omega})
\\ && \times
F\left(-{x\over 2}+{Q-\alpha_1\over 2},{x\over 2}+{Q-\alpha_2\over 2};{Q-\alpha_1\over 2}+{x\over 2},-{x\over 2}+{Q-\alpha_2\over 2};{\beta_1-\beta_2\over 2}\right),
\label{dualf} \eea
where
\be
\nu_1+\mu_1=\alpha_1,\quad \nu_2+\mu_2=\alpha_2,\quad \nu_1-\mu_1=\beta_1,\quad \nu_2-\mu_2=\beta_2.
\ee
Another useful representation of the $F$-function was derived in \cite{Kashaev}  (see also \cite{BSS})
\begin{eqnarray}\nonumber &&
F(\nu_1,\nu_2;\mu_1,\mu_2;x)=\gamma^{(2)}(\alpha_1,\alpha_2;\mathbf{\omega})
\gamma^{(2)} \Bigl(\pm x+Q-{\alpha_1+\alpha_2\over 2};\mathbf{\omega}\Bigr)
\\ \label{find1} && \makebox[2em]{}
\times\gamma^{(2)}(\nu_1+\mu_2,\nu_2+\mu_1;\mathbf{\omega}) \,
e^{{\pi\textup{i} \over \omega_1\omega_2} x(\mu_1-\nu_2)}
\\ \nonumber && \makebox[2em]{} \times
F\Big({Q-\nu_1-\mu_2\over 2},{Q+\nu_1-\mu_2\over 2}-\nu_2;
{Q-\nu_1-\mu_2\over 2},{Q-\nu_1+\mu_2\over 2}-\mu_1;x\Big).
\end{eqnarray}

After restricting the number of independent parameters by the choice
\bea\label{ruijparam}
\mu_1=g/2+\lambda_1,\quad \mu_2=g/2+\lambda_2,\quad
\nu_1=g/2-\lambda_1,\quad \nu_2=g/2-\lambda_2,
\eea
we obtain in \eqref{bd} $B=g,\, C=\lambda_1+\lambda_2$,
and equation \eqref{888} takes the form
\bea\nonumber &&
e^{-{\pi\textup{i} \over \omega_1}(\lambda_1+\lambda_2)}
{\sin{\pi\over \omega_1}(x+g)\over \sin{\pi x \over \omega_1}}F(g;\underline{\lambda};x+\omega_2)
+e^{{\pi\textup{i} \over \omega_1}(\lambda_1+\lambda_2)}{\sin{\pi\over \omega_1}(x-g)\over
\sin{\pi x \over \omega_1}}F(g;\underline{\lambda};x-\omega_2)
\\ \nonumber && \makebox[2em]{}
=2\cos{\pi\over \omega_1}(\lambda_2-\lambda_1)F(g;\underline{\lambda};x).
\eea
Representing the argument $x$ as a difference of two variables $x=x_2-x_1$ we obtain
\bea\nonumber &&
{\sin{\pi\over \omega_1}(x_2-x_1+g)\over \sin{(x_2-x_1)\pi\over \omega_1}}
F(g;\underline{\lambda};x_2-(x_1-\omega_2))
\\ \nonumber && \makebox[2em]{}
+e^{{2\pi\textup{i} \over \omega_1}(\lambda_1+\lambda_2)}
{\sin{\pi\over \omega_1}(x_2-x_1-g)\over \sin{(x_2-x_1)\pi\over \omega_1}}
F(g;\underline{\lambda};x_2-\omega_2-x_1)
\\ \nonumber && \makebox[4em]{}
=2e^{{\pi\textup{i} \over \omega_1}(\lambda_1+\lambda_2)}\cos{\pi\over \omega_1}(\lambda_2-\lambda_1)F(g;\underline{\lambda};x_2-x_1).
\eea
Noting the simple identity
\be
2e^{{\pi\textup{i} \over \omega_1}(\lambda_1+\lambda_2)}\cos{\pi\over \omega_1}(\lambda_2-\lambda_1)=e^{{2\pi\textup{i} \over \omega_1}\lambda_1}+e^{{2\pi\textup{i} \over \omega_1}\lambda_2}
\ee
and defining the function
\be\label{frh}
\Phi^g_{\lambda_1,\lambda_2}(x_1,x_2)= e^{-{2\pi\textup{i} \over \omega_1\omega_2}x_2(\lambda_1+\lambda_2)}F(g;\underline{\lambda};x_2-x_1),
\ee
we can write
\bea\label{diffig} &&
{\sin{\pi\over \omega_1}(x_2-x_1+g)\over \sin{(x_2-x_1)\pi\over \omega_1}}\Phi^g_{\lambda_1,\lambda_2}(x_1-\omega_2,x_2)+
{\sin{\pi\over \omega_1}(x_2-x_1-g)\over \sin{(x_2-x_1)\pi\over \omega_1}}\Phi^g_{\lambda_1,\lambda_2}(x_1,x_2-\omega_2)
\nonumber\\ &&  \makebox[2em]{}
=(e^{{2\pi\textup{i} \over \omega_1}\lambda_1}+e^{{2\pi\textup{i} \over \omega_1}\lambda_2})\Phi^g_{\lambda_1,\lambda_2}(x_1,x_2).
\eea
Since $\Phi^g_{\lambda_1,\lambda_2}(x_1,x_2)$ is symmetric with respect to the interchange $\omega_1 \leftrightarrow \omega_2$, we also have
\bea\label{diffig2} &&
{\sin{\pi\over \omega_2}(x_2-x_1+g)\over \sin{(x_2-x_1)\pi\over \omega_2}}\Phi^g_{\lambda_1,\lambda_2}(x_1-\omega_1,x_2)+
{\sin{\pi\over \omega_2}(x_2-x_1-g)\over \sin{(x_2-x_1)\pi\over \omega_2}}\Phi^g_{\lambda_1,\lambda_2}(x_1,x_2-\omega_1)
\nonumber \\ && \makebox[2em]{}
=(e^{{2\pi\textup{i} \over \omega_2}\lambda_1}+e^{{2\pi\textup{i} \over \omega_2}\lambda_2})\Phi^g_{\lambda_1,\lambda_2}(x_1,x_2).
\eea
From the definitions~\eqref{frh} and \eqref{newf} we find
\begin{align} \label{frh'}
	\Phi^g_{\lambda_1,\lambda_2}(x_1,x_2)={e^{-{2\pi\textup{i} \over \omega_1\omega_2}x_2(\lambda_1+\lambda_2)}\over \gamma^{(2)}(g;\mathbf{\omega})}
	\int_{-\textup{i}\infty}^{\textup{i}\infty}e^{{2\pi\textup{i} \over \omega_1\omega_2} (x_2-x_1)z}\prod_{j=1,2}\gamma^{(2)}(\tfrac{1}{2} g \pm (\lambda_j-z);\mathbf{\omega})
	{dz\over \textup{i}\sqrt{\omega_1\omega_2}},
\end{align}
which coincides with the two-particle wave function described in \cite{halik}.

The dual form \eqref{dualf} of the function $F(\underline{\nu};\underline{\mu};x_2-x_1)$
with the parametrization \eqref{ruijparam} leads to the following symmetry \cite{R11,halnas}
\bea\label{chi} &&
\Phi^g_{\lambda_1,\lambda_2}(x_1,x_2)= \Phi^{Q-g}_{x_1,x_2}(\lambda_1,\lambda_2).
\eea
  Due to this symmetry we have the second integral representation
\begin{align}
	\Phi^g_{\lambda_1, \lambda_2}(x_1, x_2) = {e^{-{2\pi\textup{i} \over \omega_1\omega_2} \lambda_2 (x_1+x_2)}\over \gamma^{(2)}( g^* ;\mathbf{\omega})}
	\int_{-\textup{i}\infty}^{\textup{i}\infty}e^{{2\pi\textup{i} \over \omega_1\omega_2} (\lambda_2-\lambda_1)z}\prod_{j=1,2}\gamma^{(2)}(\tfrac12 g^* \pm (x_j-z);\mathbf{\omega})
	{dz\over \textup{i}\sqrt{\omega_1\omega_2}},
	\label{Phi-dual}
\end{align}
where we denoted the reflected coupling constant
\begin{align}
	 g^*  = Q - g = \omega_1 + \omega_2 - g.
\end{align}

Using the transformation \eqref{find1} with the restricted parameters \eqref{ruijparam} we derive the reflection formula \cite{Belousov:2023znq}
\bea\label{sabo2}
\Phi^g_{\lambda_1,\lambda_2}(x_1,x_2)=\gamma^{(2)}(Q-g\pm x_d,
g\pm \lambda_d;\mathbf{\omega})\Phi^{Q-g}_{\lambda_1,\lambda_2}(x_1,x_2),
\eea
where $x_d=x_2-x_1$ and $\lambda_d=\lambda_2-\lambda_1$.
Recalling \eqref{chi}, we can also write
\be\label{sabo3}
\Phi^{Q-g}_{x_1,x_2}(\lambda_1,\lambda_2)=\gamma^{(2)}(Q-g\pm x_d,
g\pm \lambda_d;\mathbf{\omega})\Phi^{g}_{x_1,x_2}(\lambda_1,\lambda_2).
 \ee
Alternatively, the reflection relation can be derived by a reduction of symmetries of
the $J_h$-function \eqref{jhh}. The following transformation symmetry for the function
\eqref{jhh} was obtained in \cite{BRS} (see also \cite{Apresyan:2022erh})
\bea\label{ide1b} &&
J_h(\underline{\mu},\underline{\nu})=\prod_{ j, k =1}^2\gamma^{(2)}(\mu_j+\nu_k;\mathbf{\omega})\prod_{ j, k =3}^4\gamma^{(2)}(\mu_j+\nu_k;\mathbf{\omega})
\\ \nonumber && \makebox[2em]{} \times
J_h(\mu_1+\eta,\mu_2+\eta,\mu_3-\eta,\mu_4-\eta,\nu_1+\eta,\nu_2+\eta,\nu_3-\eta,\nu_4-\eta),
\eea
where  $\eta={1\over 2}(Q-\mu_1-\mu_2-\nu_1-\nu_2).$
Let us set in \eqref{ide1b}:
\be\label{pr113}
\begin{aligned}
&\nu_1=\xi-x_d/2+Q/2\, ,\\
&\nu_2=-\xi-g+x_d/2+Q/2\, ,\\
&\nu_3=g/2-\lambda_1\, ,\\
&\nu_4=g/2-\lambda_2,
\end{aligned}\qquad
\begin{aligned}
&\mu_1=\xi+x_d/2+Q/2\, ,\\
&\mu_2=-\xi-x_d/2-g+Q/2\, ,\\
&\mu_3=g/2+\lambda_1\, ,\\
&\mu_4=g/2+\lambda_2\, .
\end{aligned}
\ee
It is easy to see that this parametrization satisfies the balancing condition \eqref{ball}.
With this notation one has $\eta=g-Q/2$ and the limit $\xi\to \textup{i}\infty$ in \eqref{ide1b}
yields the identity \eqref{sabo2}.

For the purpose of generalizing the derived results to $N$-particle system it is natural
to work with the function \eqref{frh}. However, in order to grasp properties of the two-particle
system sometimes it is more convenient to work in the center of mass frame, which corresponds  to setting in
\eqref{frh} $ x_2=-x_1=x/2$ and $\lambda_2=-\lambda_1=\lambda/2$:
\begin{equation}\label{Fg}
F^g_{\lambda}(x):=
\frac{1}{\gamma^{(2)}(g;\mathbf{\omega})}\int_{-\imath \infty}^{\imath \infty}
e^{{2\pi\imath \over \omega_1\omega_2} zx}
\gamma^{(2)}(\tfrac12 g\pm \tfrac12\lambda \pm z;\mathbf{\omega})
{dz\over \imath \sqrt{\omega_1\omega_2}},
\end{equation}
where two symbols $\pm$ in the $\gamma^{(2)}$-function argument mean the product of four gamma functions with all possible signs.
This function has the symmetries $F^g_{\lambda}(-x)=F^g_{\lambda}(x)$ and
\begin{equation}\label{Fg_dual}
F^g_{\lambda}(x)=F^{Q-g}_{x}(\lambda).
\end{equation}
Besides, it satisfies the equation
 \bea\label{difcm}
{\sin{\pi\over \omega_1}(x+g)\over \sin{\pi x \over \omega_1}}F^g_{\lambda}(x+\omega_2)
+ {\sin{\pi\over \omega_1}(x-g)\over \sin{\pi x \over \omega_1}}F^g_{\lambda}(x-\omega_2)
=2\cos{\pi \lambda\over \omega_1}F^g_{\lambda}(x),
\eea
or in other words
$$
L_xF^g_{\lambda}(x)=2\cos{\pi \lambda\over \omega_1} F^g_{\lambda}(x), \qquad
L_x=\frac{\sin\frac{\pi}{\omega_1}(x+g)}{\sin\frac{\pi x}{\omega_1}}e^{\omega_2\partial_x}
+\frac{\sin\frac{\pi}{\omega_1}(x-g)}{\sin\frac{\pi x}{\omega_1}}e^{-\omega_2\partial_x}.
$$

The derived second order finite-difference equation has two linearly independent solutions defined up to the multiplication by $\omega_2$-periodic functions. If one rewrites it as a $q$-difference equation and looks
for solutions in terms of $\omega_1$-periodic exponential functions, then one comes
to ordinary $q$-hypergeometric functions with $|q|<1$,
which are not symmetric in $\omega_1, \,\omega_2$. The hyperbolic regime which we are interested in
admits $|q|=1$ and in this case physical solutions of \eqref{difcm} are
$x\to -x$ symmetric and $\omega_1\leftrightarrow\omega_2$ invariant (in $2d$ CFT it
corresponds to the $b\to b^{-1}$ symmetry) given by the function \eqref{Fg}.
After replacing $x$ by $\textup{i} x$, the operator $L_x$ becomes exactly the Hamiltonian
$H_h'$ \eqref{Hh2}, $L_{\textup{i} x}=H_h'$. Because of the symmetry \eqref{Fg_dual},
the variable $\lambda$ should be replaced by $\textup{i} \lambda$ as well. As a result,
we obtain physical solutions of the eigenvalues problems
$$
H_hF^g_{\textup{i}\lambda}(\textup{i}x)=EF^g_{\textup{i}\lambda}(\textup{i}x), \quad
H_h'F^g_{\textup{i}\lambda}(\textup{i}x)=E'F^g_{\textup{i}\lambda}(\textup{i}x), \quad
E= 2\ch{\pi \lambda\over \omega_2}, \quad E'=2\ch{\pi \lambda\over \omega_1}.
$$
For real $\omega_{1,2}$ and $\lambda$
 eigenvalues of both self-adjoint Hamiltonians $H_h$ and $H_h'$ are real and bounded from below, $E,E'\geq 2$.
However, for $\omega_2=\bar\omega_1$ the spectrum of the self-adjoint operator
$H_h+H_h'$ is not bounded from below
$$
E+E'=4\ch\frac{\pi \lambda\Re(\omega_1)}{|\omega_1|^2}
\cos\frac{\pi \lambda \Im (\omega_1)}{|\omega_1|^2}\in\mathbb{R}.
$$

\section{Complex degeneration of the wave function} \label{sec:comp-func}

  Let us derive the $\omega_1 + \omega_2 \to 0$ limiting form of the hyperbolic wave function $\Phi^g_{\lambda_1, \lambda_2}(x_1, x_2)$ discussed in the previous section. Our starting point is the integral representation~\eqref{Phi-dual}
\begin{align}
	\Phi^g_{\lambda_1, \lambda_2}(x_1, x_2) = {e^{-{2\pi\textup{i} \over \omega_1\omega_2} \lambda_2 (x_1+x_2)}\over \gamma^{(2)}( g^* ;\mathbf{\omega})}
	\int_{-\textup{i}\infty}^{\textup{i}\infty}e^{{2\pi\textup{i} \over \omega_1\omega_2} (\lambda_2-\lambda_1)z}\prod_{j=1,2}\gamma^{(2)} \bigl(\tfrac12 g^* \pm (x_j-z);\mathbf{\omega} \bigr)
	{dz\over \textup{i}\sqrt{\omega_1\omega_2}},
	\label{frhd}
\end{align}
  where $ g^*  =\omega_1 + \omega_2 -g$. Here we assume that $\Re \omega_j > 0$ and
\begin{align} \label{Phi-assump}
	| \Re x_j | < \tfrac{1}{2} \Re  g^* , \qquad |\Re(\lambda_2 - \lambda_1) | < \Re g.
\end{align}
The first condition ensures that two sets of the integrand poles
\begin{align}\label{Phi-poles}
	z_{\mathrm{poles}}^{\pm} = x_j \pm \bigl( \tfrac{1}{2} g^*  + m_1 \omega_1 + m_2 \omega_2 \bigr), \qquad m_1, m_2 \in \mathbb{Z}_{\geq 0},
\end{align}
are separated by the integration contour. Under the second condition the integral is absolutely convergent due to
asymptotics of the integrand gamma functions~\eqref{gamasym},~\eqref{gamasym2}.

Now, as in Section~\ref{sec:cr-deg}, we parametrize periods
\begin{align}
	\omega_1 = \imath + \delta, \qquad \omega_2 = - \imath  + \delta
\end{align}
and consider the limit $\delta \to 0^+$. Then the integrand poles~\eqref{Phi-poles} pinch the
integration contour, since $|\Re x_j| < \Re  g^* /2 < \delta$. To avoid this pinching we follow the
strategy suggested in \cite{Sarkissian:2020ipg}.

Recall the degeneration of the hyperbolic gamma function to the complex gamma function~\eqref{lim1}
\be\label{gam2lim2}
\gamma^{(2)}(\textup{i}\sqrt{\omega_1\omega_2}(n+u\delta);\mathbf{\omega})
\stackreb{=}{\delta\to 0^+} e^{\frac{\pi \textup{i}}{2}n^2} (4\pi\delta)^{\textup{i}u-1}{\bf \Gamma}(u,n),
\qquad \sqrt{\frac{\omega_1}{\omega_2}}=\textup{i}+\delta,
\ee
where $n\in \mathbb{Z}, \, u\in\mathbb{C}$.
Let us substitute in \eqref{frhd} the parametrizations
\begin{align}\label{zbi1new}
\lambda_j=\textup{i}\sqrt{\omega_1\omega_2}(N_j+\beta_j),\qquad
 x_j=\textup{i}\sqrt{\omega_1\omega_2}(n_j+u_j\delta), \qquad g^* =\textup{i}\sqrt{\omega_1\omega_2}(r+h\delta),
\end{align}
where $\beta_j \in \mathbb{R}$,  $u_j, h\in\mathbb{C}$,   $N_j \in 2\mathbb{Z}$, $r \in \mathbb{Z}$
and $n_1, n_2$ are taken to be simultaneously both integer or half-integer.
  We take $N_j$ to be even integers in order to avoid discussion of the technical details
related to the sign alternating factor $(-1)^{Nm}$ for odd $N$ in the asymptotic formula \eqref{lim2'},
as described in \cite{BSS}.
Then conditions~\eqref{Phi-assump} hold for small enough~$\delta$ if
\begin{equation} \label{uh-cond}
	| \Im  u_j | < -\tfrac{1}{2} \Im  h < 1.
\end{equation}
Furthermore, in the limit $\delta \to 0^+$ integrand poles~\eqref{Phi-poles} approach the points $\imath \mathbb{Z}$ or $\imath (\mathbb{Z} + 1/2)$ for $n_j + r/2$ integer or half-integer correspondingly. Hence, we split the integral~\eqref{frhd} into an infinite sum
\begin{align}\nonumber
		\int_{-\imath \infty}^{\imath \infty} f(z) \, \frac{dz}{\imath \sqrt{\omega_1 \omega_2}} = \int_{- \infty}^{ \infty} f(\imath \sqrt{\omega_1 \omega_2} \, \tilde{z}) \, d\tilde{z} & = \sum_{k \in \mathbb{Z} + \epsilon} \int_{k - 1/2}^{k + 1/2} f(\imath \sqrt{\omega_1 \omega_2} \, \tilde{z}) \, d\tilde{z} \\[6pt]
		& = \delta \sum_{k \in \mathbb{Z} + \epsilon} \int_{- 1/2\delta}^{1/2\delta} f( \imath \sqrt{\omega_1 \omega_2} (k + y \delta)) \, dy, \label{int-to-sum}
\end{align}
where $\epsilon \in \{0, \tfrac{1}{2} \}$ is chosen so that $n_j + r/2 + \epsilon \in \mathbb{Z}$. In total, in the above steps we changed the integration variable $z \to y$, where
\begin{align}
	z=\imath \sqrt{\omega_1\omega_2}(k+y\delta).
\end{align}

Initially the distance between integrand poles~\eqref{Phi-poles} and integration contour $\imath \mathbb{R}$ is proportional to $\delta$.
However, due to the rescaling of the integration variable by $\delta$ there is no contour pinching in the integrals from the last
line in~\eqref{int-to-sum}. Instead, as $\delta \to 0^+$ the integration
limits become infinite, and in the integrand we can apply the limiting formula~\eqref{gam2lim2}
\begin{align}
	\begin{aligned}
		\gamma^{(2)} \bigl(\tfrac12 g^* \pm (x_j-z);\mathbf{\omega} \bigr) & \underset{\delta \to 0^+}{=} e^{ \frac{\pi \imath}{2}  [r/2 + (n_j - k)]^2 + \frac{\pi \imath}{2}  [r/2 - (n_j - k) ]^2  } \, (4 \pi \delta)^{\imath h - 2} \\[6pt]
		& \;\; \times {\bf \Gamma}\bigl(\tfrac12 h \pm(u_j-y),\tfrac12 r\pm (n_j-k) \bigr).
	\end{aligned}
\end{align}
Recall that in the taken notation we have
\begin{align}
	{\bf \Gamma}(a \pm b, n \pm m) = {\bf \Gamma}(a + b, n + m) \, {\bf \Gamma}(a - b, n - m).
\end{align}
A rigorous justification of the above procedure (including uniformity of the bounds for the integrand, which allows pulling the limit inside the integral) is discussed in~\cite{Sarkissian:2020ipg}.

Under the parametrizations~\eqref{zbi1new} the exponential function in the integrand~\eqref{frhd} reads
\begin{align}
	e^{\frac{2 \pi \imath}{\omega_1 \omega_2} (\lambda_2 - \lambda_1) z} = e^{2\pi \imath ((\beta_1 - \beta_2) k + (N_1 - N_2) y \delta + (\beta_1 - \beta_2) y \delta)}.
\end{align}
Take the limit $\delta \to 0^+$ simultaneously with $|N_j| \to \infty$ in such a way
that $N_j \delta \to \alpha_j \in \mathbb{R}$, where $\alpha_j$ are two arbitrary fixed numbers.
This can be achieved by taking $N_j=[\alpha_j/\delta]_e$ -- the biggest even integer smaller than
$\alpha_j/\delta$. Since $\alpha_j/\delta-N_j$ is a bounded quantity,
for $\delta\to 0^+$ one has $\alpha_j -N_j \delta\to 0$.
As a result, we obtain
\begin{equation}
	e^{\frac{2 \pi \imath}{\omega_1 \omega_2} (\lambda_2 - \lambda_1) z} \underset{\substack{\delta \to 0^+ \\[2pt] |N_j| \to \infty} }{=} e^{2\pi \imath ((\beta_1 - \beta_2) k + (\alpha_1 - \alpha_2) y)}.
\end{equation}
In the same spirit the prefactor in front of the integral in~\eqref{frhd} has the limit
\begin{align}
	{e^{-{2\pi\textup{i} \over \omega_1\omega_2} \lambda_2 (x_1+x_2)}\over \gamma^{(2)}( g^* ;\mathbf{\omega})} \underset{\substack{\delta \to 0^+ \\[2pt] |N_j| \to \infty} }{=} e^{-\frac{\pi \imath}{2} r^2} \, (4\pi \delta)^{-\imath h + 1} \, \frac{e^{2\pi\textup{i}  (\beta_2(n_1+n_2)+\alpha_2(u_1+u_2))}}{ {\bf \Gamma} (h,r) }.
\end{align}

Collecting all above formulas together we obtain the limiting form of the wave function
\be\label{complim}
\Phi^{g}_{\lambda_1,\lambda_2}(x_1,x_2) \underset{\substack{\delta \to 0^+ \\[2pt] |N_j| \to \infty} }{=} e^{\pi\textup{i}  \Lambda(n_1,n_2,r)}(4\pi\delta)^{\imath h-2} \, F^{r,h}_{\underline{\alpha},\underline{\beta}}(\underline{u},\underline{n}),
\ee
where we denote
\bea\label{disfu} && \makebox[-1em]{}
F^{r,h}_{\underline{\alpha},\underline{\beta}}(\underline{u},\underline{n})={e^{2\pi\textup{i}  (\beta_2(n_1+n_2)+\alpha_2(u_1+u_2))} \over {4\pi\bf \Gamma}(h,r)} \\ \nonumber
&&\times\sum_{k\in \mathbb{Z}+\epsilon}\int_{-\infty}^{\infty}e^{2\pi \imath ((\beta_1 - \beta_2) k + (\alpha_1 - \alpha_2) y)}
\prod_{j=1,2}{\bf \Gamma}\bigl(\tfrac12 h \pm(u_j-y),\tfrac12 r\pm (n_j-k) \bigr) \, dy,
\eea
with $\epsilon \in \{0, \tfrac{1}{2} \}$ fixed from the condition $n_j + \tfrac{1}{2}r+ \epsilon \in \mathbb{Z}$, and
\begin{align} \label{Lambda}
	\Lambda(n_1,n_2,r)=
	\left\{
	\begin{aligned}
		& (n_1+n_2)(1-r)+\tfrac{1}{2} r^2, &&  \qquad n_1,n_2 \in \mathbb{Z},\\[4pt]
		& (n_1+n_2)(1-r)+\tfrac{1}{2} r^2+r+1, && \qquad n_1,n_2 \in \mathbb{Z}+\tfrac{1}{2}.
	\end{aligned} \right.
\end{align}
  Notice that the number of coordinates in the limit is doubled
\begin{align}
	x_1, x_2 \quad \to \quad \underline{u}= (u_1, u_2), \; \; \underline{n} = (n_1, n_2).
\end{align}
This is why we take $|N_j| \to \infty$, ensuring that the number of spectral parameters also doubles $\lambda_1, \lambda_2 \to \underline{\alpha}=(\alpha_1,\alpha_2),\, \underline{\beta}=(\beta_1,\beta_2)$.

Introduce the following combinations of parameters
\begin{align} \label{not-comb}
	\begin{aligned}
		& \gamma_j = \alpha_j + \imath \beta_j, \qquad  z_j = \frac{n_j + \imath u_j}{2}, &&\qquad  \rho = \frac{r + \imath h}{2}, \qquad\quad w = \frac{k + \imath y}{2}, \\[6pt]
		& \bar{\gamma}_j = \alpha_j - \imath \beta_j, \qquad z_j' = \frac{-n_j + \imath u_j}{2}, &&\qquad  \rho' = \frac{-r + \imath h}{2}, \qquad w' = \frac{-k + \imath y}{2}.
	\end{aligned}
\end{align}
Then, using another common notation for the complex gamma function
\begin{align}
{\bf \Gamma}(u, n) = {\bf \Gamma} \biggl( \frac{n + \imath u}{2} \bigg| \frac{-n + \imath u}{2}  \biggr),
\end{align}
we can rewrite function~\eqref{disfu} in an alternative way
\begin{align}
	\begin{aligned}
		& \makebox[-1em]{} F^{r,h}_{\underline{\alpha},\underline{\beta}}(\underline{u},\underline{n}) = \frac{e^{2\pi (\gamma_2 (z_1 + z_2) + \bar{\gamma}_2(z_1' + z_2'))}}{4\pi {\bf \Gamma} (\rho | \rho')} \\[6pt]
		& \times \sum_{k \in \mathbb{Z} + \epsilon} \int_{-\infty}^{\infty} e^{2\pi ((\gamma_1 - \gamma_2) w + (\bar{\gamma}_1 - \bar{\gamma}_2)w')} \, \prod_{j = 1,2} {\bf \Gamma} \bigl( \tfrac{1}{2}\rho \pm (z_j - w) \big| \tfrac{1}{2} \rho' \pm (z_j' - w') \bigr) \, dy.
	\end{aligned}
\end{align}

  As discussed in the previous section, the hyperbolic wave function satisfies two difference equations~\eqref{diffig} and \eqref{diffig2}.
  Their complex degeneration has been already described in Section~\ref{sec:cr-deg} with a minor difference: here we consider
  functions of two coordinates $x_1,  x_2$, whereas in Section~\ref{sec:cr-deg} we work in the center of mass frame.

Applying the limit $\delta\to 0^+$ to the difference equation \eqref{diffig2}, with the help of relations
\begin{eqnarray*} &&
	{\textup{i}\over \omega_2}\sqrt{\omega_1\omega_2}(n+u\delta) = -n+\textup{i}\delta(n+\textup{i}u)+O(\delta^2),\\
	&&\textup{i}\sqrt{\omega_1\omega_2}(n+u\delta)-\omega_1
	=\textup{i}(n-1+(u+\textup{i})\delta)+O(\delta^2),
\end{eqnarray*}
we obtain
\bea \nonumber &&
{z_2-z_1+1-\rho \over z_2-z_1}F^{r,h}_{\underline{\alpha},\underline{\beta}}(u_1+\textup{i},u_2, n_1-1, n_2)+{z_2-z_1-1+\rho\over z_2-z_1}F^{r,h}_{\underline{\alpha},\underline{\beta}}(u_1,u_2+\textup{i},n_1,n_2-1)
\\[6pt] && \makebox[4em]{}
=-(e^{-2\pi \gamma_1}+e^{-2\pi \gamma_2})F^{r,h}_{\underline{\alpha},\underline{\beta}}(\underline{u},\underline{n}). \label{Feq1}
\eea
  Notice that for the variables ~\eqref{not-comb} the shifts $(u_j, n_j) \to (u_j + \imath, n_j - 1)$
are equivalent to $(z_j, z_j') \to (z_j - 1, z_j')$, that is the primed coordinates do not change.
Similarly, taking the limit $\delta\to 0^+$ in ~\eqref{diffig}, we arrive at the equation
\bea \nonumber &&
{z'_2-z'_1+1-\rho'\over z'_2-z'_1}F^{r,h}_{\underline{\alpha},\underline{\beta}}(u_1+\textup{i},u_2, n_1+1,n_2)+{z'_2-z'_1-1+\rho'\over z'_2-z'_1}F^{r,h}_{\underline{\alpha},\underline{\beta}}(u_1,u_2+\textup{i},n_1,n_2+1)
\\[6pt] && \makebox[4em]{}
=-(e^{-2\pi \bar{\gamma}_1}+e^{-2\pi \bar{\gamma}_2})F^{r,h}_{\underline{\alpha},\underline{\beta}}(\underline{u},\underline{n}). \label{Feq2}
\eea
  Again, the shifts here are equivalent to $(z_j, z_j') \to (z_j, z_j' - 1)$.

There is one neat point regarding the above difference equations. During the derivation we assume that $u, h$ satisfy the condition~\eqref{uh-cond}, that is $| \Im  u_j | < - \tfrac{1}{2} \Im  h < 1$. As shown in \cite[Section 1.4]{Neretin:2019xze}, the same condition provides conditional convergence of the integral and absolute convergence of the series in the expression~\eqref{disfu}. However, in the difference equations above we have shifts $u_j \to u_j \pm \imath$, which spoil this condition.
A possible workaround is as follows. Let us establish the above difference equations under the assumptions
\begin{equation}\label{uh-cond2}
	u_j \in \mathbb{R}, \qquad \Im  h \in (-2, -1),
\end{equation}
which are stronger than~\eqref{uh-cond}. At the end these restrictions can be relaxed using the analytic
continuation (see \cite[Proposition 1.5]{Neretin:2019xze}).

Note that under conditions~\eqref{uh-cond2} the limiting formula~\eqref{complim} for the hyperbolic wave function $\Phi^g_{\lambda_1, \lambda_2}(x_1,x_2)$ holds true. To arrive at the difference equations~\eqref{Feq1},~\eqref{Feq2} we need to establish limiting formulas for the same function with the shifts of coordinates: $\Phi^g_{\lambda_1, \lambda_2}(x_1 - \omega_j, x_2)$ and $\Phi^g_{\lambda_1, \lambda_2}(x_1, x_2 - \omega_j)$. Since it is symmetric in $x_1, x_2$ and $\omega_1, \omega_2$, it is sufficient to consider $\Phi^g_{\lambda_1, \lambda_2}(x_1 - \omega_1, x_2)$.

To analytically continue hyperbolic wave function $\Phi^g_{\lambda_1, \lambda_2}(x_1,x_2)$ in $x_1$, we deform contour in the integral representation~\eqref{frhd}, so that it separates two sets of the integrand poles~\eqref{Phi-poles}. Conditions~\eqref{uh-cond2} ensure that for the function $\Phi^g_{\lambda_1, \lambda_2}(x_1 - \omega_1,x_2)$ we can take a straight contour
\begin{equation}
	C = \imath \mathbb{R} - \tfrac{1}{2} \delta.
\end{equation}
Indeed, in this case $x_1, x_2 \in \imath \mathbb{R}$, and the poles closest to $C$ are
\begin{equation}
	z_{+} = x_1 - \omega_1 + \tfrac{1}{2}  g^* , \qquad z_{-} = x_2 - \tfrac{1}{2} g^* .
\end{equation}
They are located from different sides of the contour $C$ at the distance
\begin{align}
	\tfrac{1}{2} (\Re  g^*  - \delta) = \tfrac{1}{2} \delta (-\sqrt{1 + \delta^2} \Im  h - 1) > 0.
\end{align}
Since the contour is straight, we can do the same steps, as before, to take the limit $\delta \to 0^+$ of the function $\Phi^g_{\lambda_1, \lambda_2}(x_1 - \omega_1, x_2)$. After all we end up with the function $F^{r, h}_{\underline{\alpha}, \underline{\beta}} (u_1 + \imath, u_2, n_1 - 1, n_2)$ given by the same expression~\eqref{disfu} modulo integration contour, which now becomes $\mathbb{R} + \tfrac{\imath}{2}  $. This function enters the first difference equation~\eqref{Feq1}, and by the same procedure we can obtain all other functions from equations~\eqref{Feq1},~\eqref{Feq2}.

Now consider the hyperbolic wave function in the center of mass frame
\be
F^{g}_{\lambda}(x) = \Phi^g_{-\frac{\lambda}{2}, \frac{\lambda}{2}}\bigl( -\tfrac{x}{2}, \tfrac{x}{2} \bigr)={1\over \gamma^{(2)}( g^* ;\mathbf{\omega})}\int_{-\textup{i}\infty}^{\textup{i}\infty}
e^{{2\pi \textup{i} \over \omega_1\omega_2} \lambda z}
\gamma^{(2)} \bigl( \tfrac{1}{2}  g^*  \pm \tfrac{1}{2}x \pm z;\mathbf{\omega} \bigr) \, {dz\over \imath \sqrt{\omega_1\omega_2}}.
\label{dufun}\ee
As before, take the periods $\omega_1 = \bar{\omega}_2 = \imath + \delta$ and parametrize other variables
\be\label{zbi1new'}
\lambda=\textup{i}\sqrt{\omega_1\omega_2}(N+\beta),\qquad x=\textup{i}\sqrt{\omega_1\omega_2}(n+u\delta),\qquad
 g^* =\textup{i}\sqrt{\omega_1\omega_2}(r+h\delta),
\ee
where $\beta \in \mathbb{R}$,  $u, h\in\mathbb{C}$   such that $|\Im  u| < -\Im  h < 2$, $N \in 2\mathbb{Z}$ and $n,r \in \mathbb{Z}$. Then from the formula~\eqref{complim} we have
\begin{align} \label{complim-cm}
	F^g_{\lambda}(x) \underset{\substack{\delta \to 0^+, \, |N| \to \infty \\[2pt] N\delta \to \alpha} }{=} e^{\pi \imath (\frac{1}{2}r^2 + n(r + 1))} \, (4\pi \delta)^{\imath h - 2} \, F^{r,h}_{\alpha,\beta}(u, n)
\end{align}
where
\begin{align} \label{FCM}
	F^{r,h}_{\alpha,\beta}(u, n) = \frac{1}{4\pi {\bf \Gamma} (h,r)} \sum_{k \in \mathbb{Z} + \epsilon} \int_{-\infty}^\infty e^{-2\pi \imath (\beta k + \alpha y)} \, {\bf \Gamma}(\tfrac12 h \pm \tfrac12 u \pm y,\tfrac12 r\pm \tfrac12 n \pm k) \, dy,
\end{align}
with $\epsilon \in \{0, \tfrac{1}{2} \}$ chosen from the demand that $\tfrac12 (r+n)+\epsilon \in \mathbb{Z}$.

Notice that the above function is related to its two variable version~\eqref{disfu} by the formula
\begin{align} \label{F2-F1}
	F^{r,h}_{\underline{\alpha}, \underline{\beta}}(\underline{u}, \underline{n}) = e^{\pi \imath((\beta_1 + \beta_2)(n_1 + n_2) + (\alpha_1 + \alpha_2)(u_1 + u_2))} \, F^{r,h}_{\alpha_2 - \alpha_1, \beta_2 - \beta_1}(u_2-  u_1, n_2 - n_1),
\end{align}
which can be obtained by shifting integration and summation variables $y \to y - \tfrac{1}{2}(u_1 + u_2)$, $k \to k - \tfrac{1}{2} (n_1 + n_2)$ in the expression~\eqref{FCM}.   Consequently, defining the variables
\be \label{comb-not2}
z=\frac{n+\imath u}{2},\qquad z'=\frac{-n+\imath u}{2}, \qquad \gamma=\alpha+\imath \beta,\qquad \bar{\gamma}=\alpha-\imath \beta,
\ee
we obtain difference equations on the wave function in the center of mass frame
\begin{align}
	& {z+1-\rho\over z}F^{r,h}_{\alpha,\beta}(u-\imath ,n+1)+{z-1+\rho\over z}F^{r,h}_{\alpha,\beta}(u+\imath ,n-1)=-2\ch\pi\gamma \, F^{r,h}_{\alpha,\beta}(u,n), \\[6pt]
	& {z'+1-\rho'\over z'}F^{r,h}_{\alpha,\beta}(u-\imath ,n-1)+{z'-1+\rho'\over z'}F^{r,h}_{\alpha,\beta}(u+\imath ,n+1)=-2\ch \pi \bar{\gamma} \, F^{r,h}_{\alpha,\beta}(u,n).
\end{align}
These equations represent eigenvalue problems for the Hamiltonians \eqref{Hcr}
and \eqref{H'_cr} with $b = 1 - \rho$ and $b' = 1 - \rho'$
\begin{align}
H_{cr} \, F^{r,h}_{\alpha\beta}(u,n)=-2\ch\pi\gamma \, F^{r,h}_{\alpha\beta}(u,n),\qquad
H_{cr}' \, F^{r,h}_{\alpha\beta}(u,n)=-2\ch \pi \bar{\gamma} \, F^{r,h}_{\alpha\beta}(u,n).
\label{fd_eqs}\end{align}

\section{The dual picture} \label{sec:dual}

As discussed in Section~\ref{sec:hyp-func}, the hyperbolic wave function has the symmetry
\begin{align} \label{Phi-sym}
	\Phi^g_{\lambda_1, \lambda_2}(x_1, x_2) = \Phi^{ g^* }_{x_1, x_2}(\lambda_1, \lambda_2).
\end{align}
It implies that this function satisfies not only the difference equations~\eqref{diffig},~\eqref{diffig2} with respect to the coordinates $x_1, x_2$, but also the same equations with respect to the spectral parameters $\lambda_1, \lambda_2$ (modulo the reflection of the coupling constant $g \to  g^* $).

In the previous section we showed that for the complex degeneration $\omega_1 + \omega_2 = 2\delta \to 0^+$ of the above function one parametrizes $\lambda_j = \imath \sqrt{\omega_1 \omega_2}(N_j + \beta_j)$ and takes $|N_j| \to \infty$ so that the products $N_j \delta$ go to some fixed numbers $\alpha_j$. At the level of difference equations the same limit is considered in Section~\ref{sec:comp-cs} (in the center of mass frame). In this limit two difference equations degenerate into two differential ones, which correspond to a complex version of the hyperbolic Calogero-Sutherland model, or equivalently, to a system of two complex conjugated hypergeometric equations.

The solution of the latter hypergeometric equations admits an Euler type integral representation,
which is different from the Mellin-Barnes type representation~\eqref{disfu}.
In this section we demonstrate how to obtain this Euler representation from the Mellin-Barnes one.

As shown in \cite[Section 5]{BSS}, the following identity holds true
\begin{equation}
\frac{\bm{\Gamma}\bigl(s\pm u, l\pm k   \bigr)}{\bm{\Gamma}(2s,2l)} = 4\pi\int_{\mathcal{C}} d^2\tau
\frac{e^{2\pi\imath(\tau_1 u+\tau_2 k)}}{( 2 \ch \pi(\tau_1+\imath\tau_2))^{l+\imath s} \, ( 2 \ch \pi(\tau_1-\imath\tau_2))^{-l+\imath s}},
\label{comp_beta_mod}
\end{equation}
where $d^2\tau = d\tau_1d\tau_2$, we integrate over the cylinder $\mathcal{C}=\mathbb{R} \times [0,1]$
and assume $l \pm k \in \mathbb{Z}$. In fact, this is just a complex analogue of the Euler integral for
beta function rewritten in the exponential variables. Note that under the assumption $l \pm k \in \mathbb{Z}$
the integrand is $1$-periodic in $\tau_2$, so we can integrate over $\tau_2$ along any full period $[a, 1 + a]$.

Hence, we can write
\bea \nonumber && \makebox[-2em]{}
{\bf \Gamma}\bigl(\tfrac12 h \pm(u_j-y),\tfrac12 r\pm (n_j-k) \bigr)
\\[6pt] &&
=4\pi{\bf \Gamma}(h,r)\int_{\mathcal{C}} d^2\tau
\frac{e^{2\pi\imath(\tau_1 (u_j-y)+\tau_2 (n_j-k))}}{( 2 \ch \pi(\tau_1+\imath\tau_2))^{{r+\imath h\over 2}} \, ( 2 \ch \pi(\tau_1-\imath\tau_2))^{-r+\imath h\over 2}}.
\eea
Inserting this combination of the complex gamma functions in the integrand \eqref{disfu}
  and using the Fourier transform inversion formula
\begin{align}
	\sum_{m \in \mathbb{Z}} \int_{-\infty}^{\infty} dy \, \int_\mathcal{C} d^2 \tau \, e^{2\pi\imath (\sigma_1 - \tau_1)y + 2\pi\imath (\sigma_2 - \tau_2) m} f(\tau_1,\tau_2)  = f(\sigma_1, \sigma_2),
\end{align}
which holds for $1$-periodic in $\tau_2$ functions $f(\tau_1, \tau_2)$,
we obtain the following Euler type representation
\begin{align} \label{F-Euler}
	\begin{aligned}
		& F^{r,h}_{\underline{\alpha},\underline{\beta}}(\underline{u},\underline{n})=4\pi{\bf \Gamma}(h,r) \, e^{2\pi\imath((\alpha_1+\alpha_2) u_2+(\beta_1+\beta_2)n_2)} \\
		& \;\; \times
		\int_{\mathcal{C}} d^2\tau \,
		\frac{e^{2\pi\imath(\tau_1( u_1-u_2)+\tau_2 (n_1-n_2))}}{\prod\limits_{j=1,2}( 2 \ch \pi[(\alpha_j-\tau_1)+\imath(\beta_j-\tau_2)])^{{r+ih\over 2}} \, ( 2 \ch \pi[(\alpha_j-\tau_1)-\imath(\beta_j-\tau_2)])^{-r+ih\over 2}}.
	\end{aligned}
\end{align}
  Observe that its structure is quite similar to the Mellin-Barnes representation \eqref{disfu}
with the roles of coordinates and spectral parameters interchanged. From this representation we obtain
\be\label{niph}
F^{r,h}_{\underline{\alpha},\underline{\beta}}(\underline{u},n_1+\nu,n_2+\nu)=
e^{2\pi\imath(\beta_1+\beta_2)\nu}F^{r,h}_{\underline{\alpha},\underline{\beta}}(\underline{u},n_1,n_2).
\ee
Therefore the functions with integer and half-integer discrete variables $n_1, n_2$ differ only by a phase factor.

Using notations~\eqref{not-comb} and also defining $\tau = \tau_1 + \imath \tau_2$ we rewrite the above representation in a more compact way
\begin{align}
	\begin{aligned}
		& F^{r,h}_{\underline{\alpha},\underline{\beta}}(\underline{u},\underline{n})=4\pi{\bf \Gamma}(\rho | \rho') \, e^{2\pi((\gamma_1 + \gamma_2) z_2+(\bar{\gamma}_1+\bar{\gamma}_2)z'_2)} \\[3pt]
		& \qquad \times \int_{\mathcal{C}} d^2\tau \, \frac{e^{2\pi(\tau( z_1 - z_2)+\bar{\tau} (z'_1-z'_2))}}{\prod\limits_{j=1,2}( 2 \ch \pi(\gamma_j-\tau))^{\rho} \, ( 2 \ch \pi(\bar{\gamma}_j-\bar{\tau}))^{\rho'}}.
	\end{aligned}
\end{align}
  Note that all integrands factorize into ``holomorphic'' and ``antiholomorphic'' parts. Moreover, separately these parts are identical to those, which enter the Euler type integral representation for the standard (real) hyperbolic Calogero-Sutherland system, see~\cite[Section~4]{HR-nonrel}.
Similarly to the Mellin-Barnes representation \eqref{disfu},
the above Euler type representation can be obtained
 in a limit from the hyperbolic level, which will be discussed separately~\cite{BSS2}.

Due to the relation~\eqref{F2-F1} the wave function in the center of mass frame has the following form
\be \label{FCM2}
F^{r,h}_{\alpha,\beta}(u,n) = 4\pi {\bf \Gamma}(\rho | \rho')  \int_{\mathcal{C}} d^2\tau \,
\frac{e^{2\pi (\tau z +\bar{\tau} z')}}{ \bigl( 2 \ch \pi \bigl[\pm \tfrac{1}{2} \gamma-\tau\bigr] \bigr)^{\rho} \, \bigl( 2 \ch \pi \bigl[ \pm \tfrac{1}{2}  \bar{\gamma}-\bar{\tau} \bigr] \bigr)^{\rho'}},
\ee
where we use notations~\eqref{comb-not2}. One can see that the integrand is periodic in $\tau_2$ and for $|\Im  u| < - \Im  h$
it decays as $e^{2\pi ( | \Im  u | + \Im  h) |\tau_1|}$ when $|\tau_1| \to \infty$ guaranteeing convergence of the integral.
  The symmetry~\eqref{Phi-sym} together with the results of Section~\ref{sec:comp-cs} imply that the function~\eqref{FCM2} diagonalizes the Hamiltonians of complex Calogero-Sutherland system
$$
H_{c}=- \partial_\gamma^2 - 2\pi \rho  \coth(\pi \gamma) \,\partial_\gamma - (\pi \rho)^2,
\qquad
H'_{c}= - \partial_{\bar \gamma}^2 - 2\pi \rho'  \coth(\pi \bar \gamma) \,\partial_{\bar \gamma} - (\pi \rho')^2.
$$
Namely, we have
\begin{align}
H_{c} \, F^{r,h}_{\alpha,\beta}(u,n)=-(\pi z)^2 F^{r,h}_{\alpha,\beta}(u,n),\qquad
H_{c}' \, F^{r,h}_{\alpha,\beta}(u,n)=-(\pi z')^2 F^{r,h}_{\alpha,\beta}(u,n).
\label{d_eqs}\end{align}
  Alternatively, one can prove diagonalization directly. Consider the first Hamiltonian $H_c$, the proof for the second one is symmetric. It is straightforward to check the identity
\begin{align}
	H_c \; \frac{1}{\bigl( 2 \ch \pi \bigl[\pm \tfrac{1}{2} \gamma-\tau\bigr] \bigr)^{\rho}} = - \frac{1}{4} \partial_\tau^2 \, \frac{1}{\bigl( 2 \ch \pi \bigl[\pm \tfrac{1}{2} \gamma-\tau\bigr] \bigr)^{\rho}}.
\end{align}
Hence, acting by $H_c$ operator on the wave function~\eqref{FCM2} and using this identity, one can integrate by parts, so that $\partial_\tau^2$ acts on the exponent, yielding the corresponding eigenvalue
\begin{align}
	- \tfrac{1}{4} \partial_\tau^2 \, e^{2\pi \tau z} = - (\pi z)^2 \, e^{2\pi \tau z} .
\end{align}
The boundary terms after integration by parts vanish because the integrand is periodic in~$\tau_2$ and decays exponentially as $|\tau_1| \to \infty$.

For $\rho=\rho'\in\mathbb{R}$ we have identified our differential equations \eqref{d_eqs}
with the ones in \cite{MN} for $a_{MN}=\rho/2,\, b_{MN}=1/2 +\rho/2$.  However, our difference equations
\eqref{fd_eqs} do not match with the difference equations in \cite{MN} for this choice of parameters.
The point is that our integral representation of the wave function \eqref{FCM2} differs from
the one given in \cite{MN} and, as a result, the direct correspondence between these two
bispectralities is not seen explicitly.

So, the wave function $F^{r,h}_{\alpha,\beta}(u,n)$ solves two spectral problems corresponding to complex rational Ruijsenaars and complex hyperbolic Calogero-Sutherland models. As we argued, this fact is a degeneration of the bispectral symmetry of the hyperbolic Ruijsenaars model~\eqref{Phi-sym}. Since this symmetry also holds for the hyperbolic model with $N$ particles~\cite{Belousov:2023bisp}, we expect the same duality between complex models in the general case.

\section{Limit of the reflection symmetry}

As discussed in Section~\ref{sec:hyp-func}, the hyperbolic wave function has the following symmetry with respect to the reflection of coupling constant~\eqref{sabo3}
\be\label{sabo4}
\gamma^{(2)}( g^* \pm [\lambda_2 - \lambda_1]; \omega) \, \Phi^{g}_{\lambda_1,\lambda_2}(x_1,x_2)=\gamma^{(2)}( g^* \pm [x_2 - x_1]; \omega) \, \Phi^{ g^* }_{\lambda_1,\lambda_2}(x_1,x_2).
\ee
  In this section we derive its complex limit. As before (see~\eqref{zbi1new}), we parametrize periods $\omega_1 = \bar{\omega}_2 =\imath + \delta$ together with other variables
\begin{align} \label{var-param}
	\lambda_j=\textup{i}\sqrt{\omega_1\omega_2}(N_j+\beta_j),\qquad
	x_j=\textup{i}\sqrt{\omega_1\omega_2}(n_j+u_j\delta), \qquad g^* =\textup{i}\sqrt{\omega_1\omega_2}(r+h\delta),
\end{align}
and take $\delta \to 0^+$, $|N_j| \to \infty$, so that $N_j \delta \to \alpha_j$ for some fixed numbers $\alpha_j$.

  First, let us obtain limits of the gamma function prefactors on both sides of~\eqref{sabo4}.
  Using the reflection formula~\eqref{g-refl} we rewrite the left-hand side expression
\begin{align} \label{lhs-f}
	\gamma^{(2)}( g^* \pm [\lambda_2 - \lambda_1]; \omega) = \frac{\gamma^{(2)}( g^*  + \lambda_2 - \lambda_1)}{\gamma^{(2)}(g + \lambda_2 - \lambda_1)} = \frac{\gamma^{(2)}( g^*  + \lambda_2 - \lambda_1)}{\gamma^{(2)}(\lambda_2 - \lambda_1)} \, \frac{\gamma^{(2)}(\lambda_2 - \lambda_1)}{\gamma^{(2)}(g + \lambda_2 - \lambda_1)} .
\end{align}
Applying the limiting formula \eqref{lim2'} to the $\gamma^{(2)}$-function ratios in~\eqref{lhs-f}, we obtain
\begin{align} \nonumber
	& \gamma^{(2)}( g^* \pm [\lambda_2 - \lambda_1];\mathbf{\omega}) \\[4pt] \nonumber
	& \underset{\substack{\delta \to 0^+, \, |N_j| \to \infty \\[2pt] N_j \delta \to \alpha_j}}{=} \bigl( 2 \sh \pi(\alpha_2-\alpha_1 + \imath (\beta_2-\beta_1)) \bigr)^{r + \imath h-1} \bigl( 2 \sh \pi(\alpha_2-\alpha_1 - \imath (\beta_2-\beta_1)) \bigr)^{ -r + \imath h-1 } \\[3pt]
	&  \hspace{0.85cm} =\bigl( 2 \sh \pi(\gamma_2-\gamma_1 ) \bigr)^{2\rho-1}
	\bigl( 2 \sh \pi(\bar{\gamma}_2-\bar{\gamma}_1 ) \bigr)^{ 2\rho'-1 }, \label{lsdd}
\end{align}
where in the last line we use notations \eqref{not-comb}.
For the $\gamma^{(2)}$-function multiplier on the right-hand side of \eqref{sabo4} we apply the limiting
relation \eqref{lim1} and obtain
\begin{align}
	\gamma^{(2)}( g^* \pm [x_2 - x_1];\mathbf{\omega}) \underset{\delta \to 0^+}{=} e^{\pi \imath (r+n_2-n_1)} \, (4\pi\delta)^{2\imath h-2} \, \bm{\Gamma}(h\pm(u_2-u_1),r\pm(n_2-n_1)).
\end{align}
Finally, we have already computed the limit of the wave function~\eqref{complim}
\be\label{complim2}
\Phi^{g}_{\lambda_1,\lambda_2}(x_1,x_2) \underset{\substack{\delta \to 0^+ \\[2pt] |N_j| \to \infty} }{=} e^{\pi\textup{i}  \Lambda(n_1,n_2,r)} \, (4\pi\delta)^{\imath h-2} \, F^{r,h}_{\underline{\alpha},\underline{\beta}}(\underline{u},\underline{n}),
\ee
where $\Lambda(n_1, n_2, r)$ is defined in~\eqref{Lambda}.
Taking into account that in the parametrization~\eqref{var-param}
$$
g=\omega_1 + \omega_2- g^* =\textup{i}\sqrt{\omega_1\omega_2}(-r+\delta (-2\imath-h) + O(\delta^2))
$$
one also has
\be\label{complim3}
\Phi^{ g^* }_{\lambda_1,\lambda_2}(x_1,x_2) \underset{\substack{\delta \to 0^+ \\[2pt] |N_j| \to \infty} }{=} e^{\pi\textup{i}  \Lambda(n_1,n_2,-r)} \, (4\pi\delta)^{-\imath h} \, F^{-r,-2\imath -h}_{\underline{\alpha},\underline{\beta}}(\underline{u},\underline{n}),
\ee
where actually $e^{\pi \imath \Lambda(n_1,n_2,-r)} = e^{\pi \imath \Lambda(n_1,n_2,r)}$,
as follows from the definition~\eqref{Lambda}.

Collecting all the factors we derive the limiting form of the reflection symmetry~\eqref{sabo4}
\begin{align}
	\label{comp-refl}
	&\bigl( 2 \sh \pi(\gamma_2-\gamma_1 ) \bigr)^{2\rho-1} \, \bigl( 2 \sh \pi(\bar{\gamma}_2-\bar{\gamma}_1 ) \bigr)^{ 2\rho'-1 } \,
	F^{r,h}_{\underline{\alpha},\underline{\beta}}(\underline{u},\underline{n})
	\nonumber \\[4pt] & \makebox[2em]{}
	=e^{\pi \imath(r+n_2-n_1)} \, \bm{\Gamma}(h\pm(u_2-u_1),r\pm(n_2-n_1)) \,
	F^{-r,-2\imath-h}_{\underline{\alpha},\underline{\beta}}(\underline{u},\underline{n}).
\end{align}
  Let us remark that this formula is consistent with the equations satisfied by the wave function. Recall that $F^{r,h}_{\underline{\alpha},\underline{\beta}}(\underline{u},\underline{n})$ diagonalizes difference operators
\begin{align}
	& M(\rho) = \frac{z_2 - z_1 + 1 - \rho}{z_2 - z_1} \, e^{- \partial_{z_1}} + \frac{z_2 - z_1 - 1 + \rho}{z_2 - z_1} \, e^{- \partial_{z_2}} ,\\[4pt]
	& M'(\rho') = \frac{z'_2 - z'_1 + 1 - \rho'}{z'_2 - z'_1} \, e^{- \partial_{z'_1}} + \frac{z'_2 - z'_1 - 1 + \rho'}{z'_2 - z'_1} \, e^{- \partial_{z'_2}},
\end{align}
see equations~\eqref{Feq1},~\eqref{Feq2}.
It is straightforward to check that the gamma function multiplier on the right-hand side of  formula~\eqref{comp-refl}
\begin{align}
	\mathcal{R} = e^{\pi \imath (r+n_2-n_1)} \, \bm{\Gamma}(h\pm(u_2-u_1),r\pm(n_2-n_1))
\end{align}
intertwines the above operators with the reflected constants
\begin{align}
	M(\rho) \, \mathcal{R}= \mathcal{R}\, M(1-\rho), \qquad M'(\rho') \, \mathcal{R} = \mathcal{R} \, M'(1-\rho').
\end{align}
Similarly the ``dual'' differential operators in the spectral variables (see Section~\ref{sec:dual})
\begin{align}
	& \hat{M}(\rho) = - \partial_{\gamma_1}^2 - \partial_{\gamma_2}^2 - 2\pi \rho \coth \pi (\gamma_1 - \gamma_2) \, (\partial_{\gamma_1} - \partial_{\gamma_2}) - 2(\pi \rho)^2, \\[4pt]
	& \hat{M}'(\rho') = - \partial_{\bar{\gamma}_1}^2 - \partial_{\bar{\gamma}_2}^2 - 2\pi \rho' \coth \pi (\bar{\gamma}_1 - \bar{\gamma}_2) \, (\partial_{\bar{\gamma}_1} - \partial_{\bar{\gamma}_2}) - 2(\pi \rho')^2
\end{align}
are intertwined by the function appearing on the left-hand side of~\eqref{comp-refl}
\begin{align}
	\hat{\mathcal{R}} = \bigl( 2 \sh \pi(\gamma_2-\gamma_1 ) \bigr)^{1 - 2\rho} \, \bigl( 2 \sh \pi(\bar{\gamma}_2-\bar{\gamma}_1 ) \bigr)^{ 1 -2\rho' },
\end{align}
that is
\begin{align}
	\hat{M}(\rho) \, \hat{\mathcal{R}}= \hat{\mathcal{R}}\, \hat{M}(1-\rho), \qquad \hat{M}'(\rho') \, \hat{\mathcal{R}} = \hat{\mathcal{R}} \, \hat{M}'(1-\rho') .
\end{align}

\section{Complex limit of the $Q$-operator}

  Baxter $Q$-operators represent a commuting family of integral operators diagonalized by the wave functions. In the case of hyperbolic Ruijsenaars model they helped to establish several symmetries, orthogonality and completeness of the wave functions~\cite{Belousov:2023bisp, Belousov:2023ort, Belousov:2023znq}. In the two-particle case Baxter operators are related to the product formulas for relativistic hypergeometric functions~\cite{halikon}. In this section we derive a complex limit of these operators and establish the corresponding product formula.

The action of the two-particle hyperbolic $Q$-operator on $\phi(x_1, x_2)$ is given by the following formula \cite{Belousov:2023sat}
\begin{multline}\label{Q2ker}
		\bigl[ Q^h_2(\lambda) \, \phi \bigr]  (x_1, x_2) = \frac{1}{2 \bigl[ \gamma^{(2)}( g^* ;
\mathbf{\omega}) \bigr]^2}  \int_{-\textup{i}\infty}^{\textup{i}\infty} \int_{-\textup{i}\infty}^{\textup{i}\infty} {dy_1\over \imath \sqrt{\omega_1 \omega_2}} \,
 {dy_2\over \imath \sqrt{\omega_1 \omega_2}}\;\, e^{- \frac{2\pi \imath}{\omega_1 \omega_2} \lambda ( x_1 + x_2 - y_1 - y_2 ) } \\[4pt]
		\times \frac{\prod\limits_{j,k = 1,2}
 \gamma^{(2)} \Bigl( \pm  (x_j - y_k) + \frac{ g^* }{2};\mathbf{\omega} \Bigr)}{ \Bigl.
 \gamma^{(2)} \bigl( \pm ( y_1 - y_2), \pm ( y_1 - y_2 )+  g^* ;\mathbf{\omega}\bigr) } \; \phi(y_1, y_2).
\end{multline}
It is diagonalized by the hyperbolic wave function
\begin{align}\label{Q2}
	Q^h_2(\lambda) \, \Phi^{g}_{\lambda_1, \lambda_2}(x_1, x_2)
 = \prod_{j = 1,2} \gamma^{(2)} \biggl( \pm  (\lambda - \lambda_j) + \frac{g}{2} ;\mathbf{\omega}\biggr) \; \Phi^{g}_{\lambda_1, \lambda_2}(x_1, x_2) .
\end{align}
Now we consider the limit $\omega_1 + \omega_2 = 2\delta \to 0^+$ of the last formula. For this we take $x_j, \lambda_j, g$ in the same form as before \eqref{zbi1new}, and in addition set
\begin{align}
\lambda = \imath\sqrt{\omega_1 \omega_2} ( N + \beta ),
\quad  N \in \mathbb{Z}, \quad \beta \in \mathbb{R},
\end{align}
where $|N| \to \infty$, so that $N \delta \to \alpha \in \mathbb{R}$.
The limit of the $Q$-operator's eigenvalue is easily calculated using formula~\eqref{lim2'}
\begin{eqnarray*} && \makebox[-1em]{}
		\gamma^{(2)} \biggl( \pm  (\lambda - \lambda_j) + \frac{g}{2};\mathbf{\omega} \biggr)
\underset{\delta\to0^+}{=} (-1)^{Nr}e^{\imath \pi r^2/2}
\\ && \times
\bigl( 2 \sh \pi(\alpha - \alpha_j + \imath(\beta - \beta_j)+\imath r/2) \bigr)^{{-r-\imath h\over 2} }
\bigl( 2 \sh \pi(\alpha - \alpha_j - \imath(\beta - \beta_j)-\imath r/2) \bigr)^{ { r - \imath h\over 2}}.
\end{eqnarray*}
The limit of eigenfunction $\Phi^{g}_{\lambda_1, \lambda_2}(x_1, x_2)$ on the right-hand side of identity \eqref{Q2} was calculated in Section~\ref{sec:comp-func}, see \eqref{complim}.

Next, consider the left-hand side of identity \eqref{Q2}. To avoid pinching of the contours we pass, as in Section~\ref{sec:comp-func}, to infinite series of integrals over intervals (see~\eqref{int-to-sum}), so that the integration variables become parametrized in the following way
\begin{align}
	y_j = \imath\sqrt{\omega_1 \omega_2} (m_j + t_j \delta), \quad m_j \in \mathbb{Z} + \epsilon, \quad t_j \in \mathbb{R},
\end{align}
where $\epsilon \in \{0, \tfrac{1}{2}\}$ is chosen from the requirement $n_j + \tfrac{1}{2}r + \epsilon \in \mathbb{Z}$. To derive the limiting form of the integrand we use the degeneration formula ~\eqref{lim1}.
As a result, defining $\gamma=\alpha+\imath\beta$, $\bar\gamma=\alpha-\imath\beta$ and also using notations~\eqref{not-comb} we arrive at the identity
\begin{eqnarray} \nonumber &&
	[\tilde{Q}_2(\alpha,\beta) \, F^{r,h}_{\underline{\alpha},\underline{\beta}}](\underline{u},\underline{n}) = F^{r,h}_{\underline{\alpha},\underline{\beta}}(\underline{u},\underline{n}) \,
e^{\imath \pi (r+(r+1)(n_1+n_2))}
\\ && \makebox[1em]{} \times
\prod_{j = 1,2}
  \, \bigl( 2 \sh \pi(\gamma - \gamma_j +\imath r/2) \bigr)^{{-r-\imath h\over 2} }
\bigl( 2 \sh \pi(\bar{\gamma} - \bar{\gamma}_j -\imath r/2) \bigr)^{ { r - \imath h\over 2}},
\label{Q22}\end{eqnarray}
where the action of operator $\tilde{Q}_2(\alpha,\beta)$ is defined by the formula
\begin{eqnarray} \nonumber &&
		\bigl[ \tilde{Q}_2(\alpha,\beta) \, \phi \bigr]  (\underline{u},\underline{n}) = \frac{1}{2 \bigl[4\pi \, \bm{\Gamma}(h,r) \bigr]^2}
\sum_{m_j\in \mathbb{Z}+\epsilon}\int_{t_j\in\mathbb{R}} dt_1 \,
 dt_2\;\, e^{ 2\pi \imath( \alpha ( u_1 + u_2 - t_1 - t_2 )+\beta ( n_1 + n_2 - m_1 - m_2 ) )}
 \\ && \makebox[2em]{}
		\times \frac{e^{\imath \pi(r+1)(m_1+m_2)}\prod\limits_{j,k = 1,2} \bm{\Gamma} \Bigl( \pm  (u_j - t_k) + \frac{h}{2}, \pm  (n_j - m_k) + \frac{r}{2}\Bigr)}{ \Bigl. \bm{\Gamma} \bigl( \pm ( t_1 - t_2),\pm ( m_1 - m_2) \bigr) \,  \bm{\Gamma} \bigl( \pm ( t_1 - t_2 )+ h, \pm(m_1 - m_2) + r\bigr) } \; \phi(\underline{t},\underline{m}).
\label{Q2ker2}\end{eqnarray}
If we shift the parameter $\beta \to \beta - \tfrac{r+1}{2}$, then the same identity can be rewritten in a more compact way
\begin{eqnarray}\nonumber &&
	[Q_2(\alpha,\beta) \, F^{r,h}_{\underline{\alpha},\underline{\beta}}](\underline{u},\underline{n}) =
\prod_{j = 1,2}   \, \bigl( 2 \ch \pi(\gamma - \gamma_j ) \bigr)^{-\rho }
\bigl( 2 \ch \pi(\bar{\gamma} - \bar{\gamma}_j) \bigr)^{-\rho'}
F^{r,h}_{\underline{\alpha},\underline{\beta}}(\underline{u},\underline{n}),
\label{Q223}\end{eqnarray}
where the simplified complex $Q$-operator has the following form
\begin{multline}\label{Q2ker2'}
		\bigl[ Q_2(\alpha,\beta) \, \phi \bigr]  (\underline{u},\underline{n}) = \frac{1}{2 \bigl[4\pi \, \bm{\Gamma}(h,r) \bigr]^2}
\sum_{m_j\in \mathbb{Z}+\epsilon}\int_{t_j\in\mathbb{R}} dt_1 \,
 dt_2\;\, e^{ 2\pi \imath( \alpha ( u_1 + u_2 - t_1 - t_2 )+\beta ( n_1 + n_2 - m_1 - m_2 ) )} \\[10pt]
		\times \frac{\prod\limits_{j,k = 1,2} \bm{\Gamma} \Bigl( \pm  (u_j - t_k) + \frac{h}{2}, \pm  (n_j - m_k) + \frac{r}{2}\Bigr)}{ \Bigl. \bm{\Gamma} \bigl( \pm ( t_1 - t_2),\pm ( m_1 - m_2) \bigr) \,  \bm{\Gamma} \bigl( \pm ( t_1 - t_2 )+ h, \pm(m_1 - m_2) + r\bigr) } \; \phi(\underline{t},\underline{m}).
\end{multline}
  Notice that this operator and its eigenvalue are quite similar to the ones for usual (real) rational Ruijsenaars model~\cite[Section 3]{Belousov:2023sat}.

Observe that the Baxter operators with different parameters are diagonalized by the same family of wave functions, which is related to their commutativity. The commutativity of hyperbolic $Q$-operators
\be
\bigl[ Q^h_2(\lambda_1), Q^h_2(\lambda_2)\bigl]=0
\label{comm_hyp}\ee
is proven in~\cite{Belousov:2023qgn}.
Now let us derive the same property for their complex analogues
\begin{align} \label{qcomm}
	\bigl[ Q_2(\alpha_1,\beta_1), Q_2(\alpha_2,\beta_2)\bigl]=0.
\end{align}
The commutator relation \eqref{comm_hyp} can be written as an identity for the kernels of the products of
integral operators
\begin{multline}\label{Q2kerp}
		 \int_{-\textup{i}\infty}^{\textup{i}\infty} \int_{-\textup{i}\infty}^{\textup{i}\infty} {dy_1\over \imath \sqrt{\omega_1 \omega_2}} \,
 {dy_2\over \imath \sqrt{\omega_1 \omega_2}}\,\Big[ e^{- \frac{2\pi \imath}{\omega_1 \omega_2} \lambda_1 ( x_1 + x_2 - y_1 - y_2 ) }
 e^{- \frac{2\pi \imath}{\omega_1 \omega_2} \lambda_2 ( y_1 + y_2 - z_1 - z_2 ) }
\\[10pt]
-  e^{- \frac{2\pi \imath}{\omega_1 \omega_2} \lambda_2 ( x_1 + x_2 - y_1 - y_2 ) }
 e^{- \frac{2\pi \imath}{\omega_1 \omega_2} \lambda_1 ( y_1 + y_2 - z_1 - z_2 ) }\Big]
 \\[6pt]
		\times \frac{\prod\limits_{j,k = 1,2}
\gamma^{(2)} \Bigl( \pm  (x_j - y_k) + \frac{ g^* }{2},
 \pm  (y_j - z_k) + \frac{ g^* }{2};\mathbf{\omega} \Bigr)}
{ \Bigl. \gamma^{(2)} \bigl( \pm ( y_1 - y_2) , \pm ( y_1 - y_2 )+  g^* ;\mathbf{\omega}\bigr) }=0.
\end{multline}

As before, the complex degeneration $\omega_1 + \omega_2 = 2\delta \to 0^+$ of this identity is established by passing to infinite sums of integrals and with the help of the limiting formula~\eqref{lim1}
\begin{multline}\label{Q2ker22}
\sum_{m_j\in \mathbb{Z}+\epsilon}\int_{t_j\in\mathbb{R}} dt_1 \,
 dt_2\,\Big[ e^{ 2\pi \imath( \alpha_1 ( u_1 + u_2 - t_1 - t_2 )+\beta_1 ( n_1 + n_2 - m_1 - m_2 ) )}
 e^{ 2\pi \imath( \alpha_2 ( t_1 + t_2 - s_1 - s_2 )+\beta_2 ( m_1 + m_2 - l_1 - l_2 ) )}
 \\[10pt]
 - e^{ 2\pi \imath( \alpha_2 ( u_1 + u_2 - t_1 - t_2 )+\beta_2 ( n_1 + n_2 - m_1 - m_2 ) )}
 e^{ 2\pi \imath( \alpha_1 ( t_1 + t_2 - s_1 - s_2 )+\beta_1 ( m_1 + m_2 - l_1 - l_2 ) )}\Big]
 \\[10pt]
		\times\frac{\prod\limits_{j,k = 1,2} \bm{\Gamma} \Bigl( \pm  (u_j - t_k) + \frac{h}{2}, \pm  (n_j - m_k) + \frac{r}{2}\Bigr)\bm{\Gamma} \Bigl( \pm  (t_j - s_k) + \frac{h}{2}, \pm  (m_j - l_k) + \frac{r}{2}\Bigr)}{ \Bigl. \bm{\Gamma} \bigl( \pm ( t_1 - t_2),\pm ( m_1 - m_2) \bigr) \,  \bm{\Gamma} \bigl( \pm ( t_1 - t_2 )+ h, \pm(m_1 - m_2) + r\bigr) } =0.
\end{multline}
This is precisely the commutativity condition~\eqref{qcomm} written as an identity for the
$Q_2(\alpha,\beta)$ integral operator kernel.

At last, let us remark that for the two-particle model the diagonalization property of Baxter operators is equivalent to the product formulas for the wave functions in the center of mass frame. It is known that the hyperbolic function $F^{g}_{\lambda}(x) = \Phi^g_{-\frac{\lambda}{2}, \frac{\lambda}{2}}\bigl( -\tfrac{x}{2}, \tfrac{x}{2} \bigr)$ satisfies the following product formula \cite{halikon}
\bea \nonumber &&
4F^g_{\lambda}(x_1)F^g_{\lambda}(x_2)=\gamma^{(2)}(g;\mathbf{\omega})\int_{-\textup{i}\infty}^{\textup{i}\infty}
{\gamma^{(2)}(g\pm z;\mathbf{\omega}) \over \gamma^{(2)}(\pm z;\mathbf{\omega})}
\\ && \makebox[2em]{} \times
\gamma^{(2)}(( g^* \pm z\pm x_1\pm x_2)/2;\mathbf{\omega}) \, F^g_{\lambda}(z){dz\over \imath \sqrt{\omega_1\omega_2}}.
\label{prfor} \eea
This identity is related to equation \eqref{Q2} in the center of mass frame $x_1+x_2=\lambda_1+\lambda_2=0$ by the Fourier transformation
(the Fourier transformed eigenvalue of the $Q$-operator becomes the second hyperbolic wave function in
 the product formula).

As we established in Section~\ref{sec:comp-func}, under parametrizations~\eqref{zbi1new'}
\begin{align}
	\lambda=\textup{i}\sqrt{\omega_1\omega_2}(N+\beta),\qquad x_j=\textup{i}\sqrt{\omega_1\omega_2}(n_j+u_j\delta),\qquad
	 g^* =\textup{i}\sqrt{\omega_1\omega_2}(r+h\delta),
\end{align}
where $N \in 2\mathbb{Z}$ and $n_j, r \in \mathbb{Z}$, the hyperbolic wave function in the center of mass frame has the limit~\eqref{complim-cm}
\begin{align}
	F^g_{\lambda}(x_j) \underset{\substack{\delta \to 0^+, \, |N| \to \infty \\[2pt] N\delta \to \alpha} }{=} e^{\pi \imath (r^2/2 + n_j(r + 1))} \, (4\pi \delta)^{\imath h - 2} \, F^{r,h}_{\alpha,\beta}(u_j, n_j)
\end{align}
with $F^{r,h}_{\alpha,\beta}(u_j, n_j)$ function defined in \eqref{FCM}.
Consequently, the complex degeneration of the product formula looks as follows
\begin{eqnarray} \nonumber &&  \makebox[-2em]{}
		4F^{r,h}_{\alpha,\beta}(u_1,n_1)F^{r,h}_{\alpha,\beta}(u_2,n_2) =  {1\over 4\pi\bm{\Gamma}(h,r)}
  \\  &&
		\times \sum_{m\in \mathbb{Z}+\epsilon}\int_{\mathbb{R}} dt \, \frac{ \bm{\Gamma} \Bigl((h \pm u_1\pm u_2 \pm t )/2, (  r \pm n_1\pm n_2 )/2 \pm m\Bigr)}{ \Bigl. \bm{\Gamma} \bigl( \pm t,\pm 2m \bigr) \,  \bm{\Gamma} \bigl( \pm t+ h, \pm 2m + r\bigr) } \; F^{r,h}_{\alpha,\beta}(t,2m).
\label{Q2ker3}\end{eqnarray}
Here $\epsilon \in \{0, \tfrac{1}{2}\}$ is fixed under the condition
that $\epsilon + \tfrac{1}{2}(r + n_1 + n_2) \in \mathbb{Z}$.

\smallskip

{\bf Acknowledgements.}
The authors thank S. Derkachov and S. Khoroshkin for interesting discussions.
This study has been partially supported by the Russian Science Foundation (grant 24-21-00466).

\appendix

\section{Properties of the hyperbolic gamma function} \label{app:hgamma}

The Faddeev modular dilogarithm \cite{Fad94}, or the hyperbolic gamma function  \cite{R11}
$\gamma^{(2)}(y;\omega_1,\omega_2)$ can be defined by several means. We use the
representation
\be
\gamma^{(2)}(u;\mathbf{\omega})= \gamma^{(2)}(u;\omega_1,\omega_2):=e^{-\frac{\pi\textup{i}}{2}
B_{2,2}(u;\mathbf{\omega}) } \gamma(u;\mathbf{\omega}),
\label{HGF}\end{equation}
where $B_{2,2}$ is the second order multiple Bernoulli polynomial
$$
B_{2,2}(u;\mathbf{\omega})=\frac{1}{\omega_1\omega_2}
\left( \Bigl(u-\frac{\omega_1+\omega_2}{2} \Bigr)^2-\frac{\omega_1^2+\omega_2^2}{12}\right)
$$
and
\be
\gamma(u;\mathbf{\omega}):= \frac{(\tilde q e^{2\pi \textup{i} \frac{u}{\omega_1}};\tilde q)_\infty}
{(e^{2\pi \textup{i} \frac{u}{\omega_2}};q)_\infty}
=\exp\left(-\int_{\mathbb{R}+\textup{i}0}\frac{e^{ux}}
{(1-e^{\omega_1 x})(1-e^{\omega_2 x})}\frac{dx}{x}\right)
\label{int_rep}\end{equation}
with
$$
{ q}=e^{2\pi\textup{i}\frac{\omega_1}{\omega_2}}, \qquad \tilde { q}
=e^{-2\pi\textup{i}\frac{\omega_2}{\omega_1}},
$$ and $(a; q)_{\infty}=\prod_{k=0}^{\infty}(1-aq^k)$.
The superindex $(2)$ indicates that this is the hyperbolic gamma function of the second
order in accordance with the order of the Barnes multiple gamma function used for its
definition. Its reciprocal is known also as the double sine function, see e.g. \cite{Belousov:2023qgn}.

This function obeys the first order difference equations
\be\label{hp1}
{\gamma^{(2)}(u+\omega_1;\mathbf{\omega})\over \gamma^{(2)}(u;\mathbf{\omega})}=2\sin{\pi u\over \omega_2},\qquad
{\gamma^{(2)}(u+\omega_2;\mathbf{\omega})\over \gamma^{(2)}(u;\mathbf{\omega})}=2\sin{\pi u\over \omega_1}
\end{equation}
and has the following asymptotics
\be\label{gamasym}
\stackreb{\lim}{u\to \infty}e^{{\pi\textup{i}\over 2}B_{2,2}(u,\omega_1,\omega_2)}\gamma^{(2)}(u;\mathbf{\omega})=1,
\quad {\rm for}\; {\rm arg}\;\omega_1<{\rm arg}\; u<{\rm arg}\;\omega_2+\pi,
\end{equation}
\be\label{gamasym2}
\stackreb{\lim}{u\to \infty}e^{-{\pi\textup{i}\over 2}B_{2,2}(u,\omega_1,\omega_2)}\gamma^{(2)}(u;\mathbf{\omega})=1,
\quad {\rm for}\; {\rm arg}\;\omega_1-\pi<{\rm arg}\; u<{\rm arg}\;\omega_2.
\end{equation}
  Besides, it has poles and zeros at the points
\begin{align}
	u_{\mathrm{poles}} = - m_1 \omega_1 - m_2 \omega_2, \qquad u_{\mathrm{zeros}} = \omega_1 + \omega_2 + m_1 \omega_1 + m_2 \omega_2, \qquad m_1, m_2 \in \mathbb{Z}_{\geq 0}.
\end{align}
It also satisfies the reflection formula
\begin{align} \label{g-refl}
	\gamma^{(2)}(u; \omega) \, \gamma^{(2)}(\omega_1 + \omega_2 - u; \omega) = 1.
\end{align}

Let us describe several qualitatively different degenerations of the hyperboilc gamma function.
First, there are two well known limits to classical functions
	\begin{align}\label{r_real}
		& \gamma^{(2)}(x \omega_1; \omega) \underset{\omega_1 \to 0^+}{=} \frac{1}{\sqrt{2\pi}} \, \biggl( \frac{2\pi \omega_1}{\omega_2} \biggr)^{x - \frac{1}{2}} \, \Gamma(x), \\[10pt]
		& \frac{\gamma^{(2)}(x + g \omega_1; \omega)}{\gamma^{(2)}(x; \omega)} \underset{\omega_1 \to 0^+}{=} \biggl( 2 \sin \frac{\pi x}{\omega_2} \biggr)^g.
\label{r_real_rat}	\end{align}
They are used in the reduction of hyperbolic Ruijsenaars model to the usual (real) rational case and to the hyperbolic Calogero-Sutherland model.

The complex analogues of the these limits have substantially more complicated form.
An analogue of \eqref{r_real},  emerging in the limit $\omega_1+\omega_2\to 0$, was
heuristically considered in \cite{BMS} and rigorously derived in \cite{Sarkissian:2020ipg}. It has the form
\begin{align}\label{lim1}
	\gamma^{(2)} \bigl( \imath \sqrt{\omega_1 \omega_2} (m + u \delta); \omega \bigr) \underset{\delta \to0^+}{=} e^{\frac{\pi \imath}{2} m^2} (4\pi \delta)^{\imath u - 1} \, \bm{\Gamma}(u, m),
\quad \sqrt{\omega_1\over \omega_2}=\textup{i}+\delta,
\end{align}
where $m \in \mathbb{Z}$, $u \in \mathbb{C}$. Here $\bm{\Gamma}(u, m)$ is the gamma function over the field
of complex numbers \cite{GGR}
\begin{align}
	\bm{\Gamma}(u,m):= \frac{\Gamma \bigl( \frac{m + \imath u}{2} \bigr)}{\Gamma \bigl( 1 +  \frac{m - \imath u}{2} \bigr)}.
\end{align}
An alternative notation is
\begin{align} \label{cg-alt}
	\bm{\Gamma}(\alpha|\alpha')  =
	\frac{\Gamma( \alpha)}{\Gamma(1-\alpha')},\quad \alpha=\frac{m + \imath u}{2},\quad \alpha-\alpha' =m\in\mathbb{Z}.
\end{align}
From the reflection relation $\Gamma(x)\Gamma(1-x)=\pi/\sin\pi x$
the following  identities are obtained
\begin{equation}
{\bf \Gamma}(\alpha|\alpha') =(-1)^{\alpha-\alpha'}{\bf \Gamma}(\alpha'|\alpha), \qquad
{\bf \Gamma}(x,-n)=(-1)^n{\bf \Gamma}(x,n),
\label{reflCgamma0}\end{equation}
and
\begin{equation}
{\bf \Gamma}(\alpha|\alpha'){\bf \Gamma}(1-\alpha|1-\alpha')  =(-1)^{\alpha-\alpha'}, \qquad
{\bf \Gamma}(x,n){\bf \Gamma}(-x-2\textup{i},n)=1.
\label{reflCgamma}\end{equation}
One has also the finite-difference equations
\begin{eqnarray*} &&
{\bf \Gamma}(\alpha+1|\alpha') ={\bf \Gamma}(x-\textup{i},n+1)=\alpha{\bf \Gamma}(\alpha|\alpha'),\quad
\\ &&
{\bf \Gamma}(\alpha|\alpha'+1) ={\bf \Gamma}(x-\textup{i},n-1)=-\alpha' {\bf \Gamma}(\alpha|\alpha').
\end{eqnarray*}

A complex analogue of the relation \eqref{r_real_rat} was derived in \cite{BSS} and it has the
following form
\begin{eqnarray}\label{lim2'} &&
e^{\pi\imath Nm}\frac{\gamma^{(2)} \bigl( \imath \sqrt{\omega_1 \omega_2} \bigl[ N + \beta + m + u \delta \bigr] ; \omega \bigr)}{\gamma^{(2)} \bigl( \imath \sqrt{\omega_1 \omega_2} \bigl[ N + \beta\bigr] ; \omega \bigr)}
\\ && \makebox[2em]{}
	\underset{ \substack{ \; \delta \to 0^+,\, |N| \to \infty \\[2pt] N\delta \to \alpha } }{=} e^{\frac{\pi \imath}{2}m^2}\, \bigl( 2 \sh \pi(\alpha + \imath \beta) \bigr)^{\frac{m + \imath u}{2} } \bigl( 2 \sh \pi(\alpha - \imath \beta) \bigr)^{ \frac{- m + \imath u}{2} }
\nonumber\end{eqnarray}
where $N, m \in \mathbb{Z}$, $\alpha\in\mathbb{R}$, $\beta, u \in \mathbb{C}$.

  Another singular degeneration derived in~\cite{Sarkissian:2020ipg} emerges in the limit $\omega_1 \to \omega_2$ and has the form
\begin{align} \label{gamma-poh-limit}
	\gamma^{(2)} \bigl( \sqrt{\omega_1 \omega_2} (m + u \delta); \omega \bigr) \underset{\delta \to 0^+}{=} e^{- \frac{\pi \imath}{2}(m - 1)^2} \, (4\pi \delta)^{m - 1} \,
\Big(1 - \frac{m + \imath u}{2}\Big)_{m-1} ,
\qquad \sqrt{\frac{\omega_1}{\omega_2}} = 1 + \imath \delta,
\end{align}
where $m \in \mathbb{Z}$, $u \in \mathbb{C}$. In this case on the right-hand side one has
the standard Pochhammer symbol
\[
(a)_m=\frac{\Gamma(a+m)}{\Gamma(a)}=
\begin{cases}
 a(a+1)\cdots(a+m-1), & \text{for} \ m>0,\quad \\
\dfrac{1}{(a-1)(a-2)\cdots(a+m)}, &\text{for} \ m<0.
\end{cases}
\]

\end{document}